\documentclass[a4paper,12pt, epsfig]{article}\def\letter{0}\def\pr{0}
\usepackage{epsfig}
\usepackage{epstopdf}
\usepackage{graphicx}
\usepackage{ifthen}
\usepackage[hypertexnames=false]{hyperref}
\hypersetup{
    colorlinks=true,
    linkcolor=blue,
    filecolor=magenta,      
    urlcolor=cyan,
    pdftitle={Overleaf Example},
    pdfpagemode=FullScreen,
    }

\usepackage{amsmath}
\usepackage[psamsfonts]{amssymb}
\usepackage{euscript}
\usepackage{latexsym}
\usepackage[arrow,matrix,curve]{xy}

\usepackage{xcolor}
\newskip\humongous \humongous=0pt plus 1000pt minus 1000pt

\newif\ifdtup

\def\Xint#1{\mathchoice
   {\XXint\displaystyle\textstyle{#1}}%
   {\XXint\textstyle\scriptstyle{#1}}%
   {\XXint\scriptstyle\scriptscriptstyle{#1}}%
   {\XXint\scriptscriptstyle\scriptscriptstyle{#1}}%
   \!\int}
\def\XXint#1#2#3{{\setbox0=\hbox{$#1{#2#3}{\int}$}
     \vcenter{\hbox{$#2#3$}}\kern-.5\wd0}}

\def\dashint{\Xint-}
\def\,{\hspace{-.1cm}}
\def\hsp{,\hspace{.7cm}}

\def\tf {\tilde{f}}

\def\fc#1#2 {\frac{n}{q}#1\frac{n}{q}#2}

\def\blu#1{\textcolor{blue}{Jarah: #1}}

\def\litwo{{\operatorname{li}_2}}

\newcommand{\vac}{\ensuremath{|0\rangle}}

\renewcommand{\sin}{\textrm{sin}}

\renewcommand{\sinh}{\textrm{sinh}}
\renewcommand{\cosh}{\textrm{cosh}}
\renewcommand{\tanh}{\textrm{tanh}}
\newcommand{\sech}{\textrm{sech}}
\newcommand{\csch}{\textrm{csch}}

\def\exp#1{\hbox{\rm exp}\left(#1\right)}

\renewcommand{\theequation}{\arabic{section}.\arabic{equation}}
\renewcommand{\(}{\begin{equation}}
\renewcommand{\)}{end{equation} \vspace{-.05in}\linebreak}

\newcounter{saveeqn}
\newcounter{savealpheqn}

\newcommand{\alpheqn}{\setcounter{saveeqn}{\value{equation}}%
  \stepcounter{saveeqn}\setcounter{equation}{0}%
  \renewcommand{\theequation}{\mbox{\arabic{section}.\arabic{saveeqn}
\alph{equation}}}
  \renewcommand{\)}{\end{equation}}}
\def\part#1{\frac{\partial}{\partial{#1}}}%
\def\group#1{\refstepcounter{equation}\setcounter{saveeqn}
 {\value{equation}}%
  \label{#1}\setcounter{equation}{0}%
\renewcommand{\theequation}{\mbox{\arabic{section}.\arabic{saveeqn}
\alph{equation}}}
  \renewcommand{\)}{\end{equation}}}
\newcommand{\reseteqn}{\setcounter{equation}{\value{saveeqn}}%
  \renewcommand{\theequation}{\arabic{section}.\arabic{equation}}%
  \renewcommand{\)}{\end{equation}}}

\newcommand{\aalpheqn}{\setcounter{saveeqn}{\value{equation}}%
  \stepcounter{saveeqn}\setcounter{equation}{0}%
  \renewcommand{\theequation}{\mbox{
        \Alph{subsection}.\arabic{saveeqn}\alph{equation}}}
   \renewcommand{\)}{\end{equation}}}
\newcommand{\areseteqn}{\setcounter{equation}{\value{saveeqn}}%
  \renewcommand{\theequation}{\Alph{subsection}.\arabic{equation}}%
  \renewcommand{\)}{\end{equation}}}

\renewcommand{\thefootnote}{\alph{footnote}}
\renewcommand{\(}{\begin{equation}}
\renewcommand{\)}{\end{equation}}
\newcommand{\ba}{\begin{eqnarray}}
\newcommand{\ea}{\end{eqnarray}}
\newcommand{\cbp}{\mathop{\vtop{\ialign{##\crcr
   $\hfil\displaystyle{}\hfil$\crcr\noalign{\kern-13pt\nointerlineskip}
   \BIG{)}\hskip0pt\crcr\noalign{\kern3pt}}}}}
\newcommand{\pa}{\mathop{\vtop{\ialign{##\crcr

$\hfil\displaystyle{\oplus}\hfil$\crcr\noalign{\kern+1pt\nointerlineskip
}
   \hspace{.08in}$^{\alpha=0}$\hskip6pt\crcr\noalign{\kern3pt}}}}}
\renewcommand{\hsp}{,\hspace{.3in}}
\newcommand{\p}{^\prime}
\newcommand{\pp}{^{\prime\prime}}

\catcode`\@=11
\def\vereq#1#2{\lower3pt\vbox{\baselineskip1.5pt \lineskip1.5pt
\ialign{$\m@th#1\hfill##\hfil$\crcr#2\crcr\sim\crcr}}}
\catcode`\@=12

\renewcommand{\(}{\begin{equation}}
\renewcommand{\)}{\end{equation}}

\def\lpin#1{\int^\Lambda_{-\Lambda} \frac{d#1}{2\pi}}
\def\pin#1{\int \frac{d#1}{2\pi}}
\def\pink#1{\int \frac{d^{#1}k}{(2\pi)^{#1}}}

\def\pinkp#1{\int \frac{d^{#1}k\p}{(2\pi)^{#1}}}

\def\Bd#1{B^\ddag_{k_{#1}}}

\def\df{\mathcal{D}_f}

\def\I{\mathcal{I}}
\renewcommand{\L}{{(\Lambda)}}

\newcommand{\beas}{\begin{eqnarray*}}
\newcommand{\eeas}{\end{eqnarray*}}

\newcommand{\bquo}{\begin{quote}}
\newcommand{\enqu}{\end{quote}}

\def\ch{{\mathcal{H}}}
\def\co{{\mathcal{O}}}
\def\tf{{\tilde{f}}}

\def\okp#1{\omega_{k\p_{#1}}}
\def\V#1{V^{(#1)}[gf(x)]}
\def\mh{\mathcal{H}}

\newcommand{\beq}{\begin{equation}}
\newcommand{\eeq}{\end{equation}}
\newcommand{\bea}{\begin{eqnarray}}
\newcommand{\eea}{\end{eqnarray}}

\newskip\humongous \humongous=0pt plus 1000pt minus 1000pt

\newif\ifdtup

\catcode`\@=11

\ifthenelse{\equal{\letter}{0}}{ %se lettere=0, commenta questo

\@addtoreset{equation}{section}
\def\theequation{\arabic{section}.\arabic{equation}}

\def\@normalsize{\@setsize\normalsize{15pt}\xiipt\@xiipt
\abovedisplayskip 14pt plus3pt minus3pt%
\belowdisplayskip \abovedisplayskip
\abovedisplayshortskip \z@ plus3pt%
\belowdisplayshortskip 7pt plus3.5pt minus0pt}

\def\small{\@setsize\small{13.6pt}\xipt\@xipt
\abovedisplayskip 13pt plus3pt minus3pt%
\belowdisplayskip \abovedisplayskip
\abovedisplayshortskip \z@ plus3pt%
\belowdisplayshortskip 7pt plus3.5pt minus0pt
\def\@listi{\parsep 4.5pt plus 2pt minus 1pt
      \itemsep \parsep
      \topsep 9pt plus 3pt minus 3pt}}

\relax

\def\section{\@startsection{section}{1}{\z@}{3.5ex plus 1ex minus  .2ex}{2.3ex plus .2ex}{\large\bf}}

\def\thesection{\arabic{section}}
\def\thesubsection{\arabic{section}.\arabic{subsection}}

\def\appendix{\setcounter{section}{0}
 \def\thesection{Appendix \Alph{section}}
 \def\thesubsection{\Alph{section}.\arabic{subsection}}
 \def\theequation{\Alph{section}.\arabic{equation}}}
\renewcommand{\theequation}{\arabic{section}.\arabic{equation}}

}{
\renewcommand{\theequation}{\arabic{equation}}

} %se letter=0, commenta questo

\begin{document}
% ========================================================================
\def\thefootnote{\fnsymbol{footnote}}
\def\thetitle{Cut-Off Kinks}
\def\autone{Jarah Evslin}
\def\auttwo{Andrew B. Royston}
\def\autthree{Baiyang Zhang}
\def\affa{Institute of Modern Physics, NanChangLu 509, Lanzhou 730000, China}
\def\affb{University of the Chinese Academy of Sciences, YuQuanLu 19A, Beijing 100049, China}
\def\affc{Department of Physics, Penn State Fayette, The Eberly Campus,
2201 University Drive, Lemont Furnace, PA 15456, USA}
\def\affd{Institute of Contemporary Mathematics, School of Mathematics and Statistics,
Henan University, Kaifeng, Henan 475004, P. R. China}

\title{Titolo}

\ifthenelse{\equal{\pr}{1}}{
\title{\thetitle}
\author{\autone}
\affiliation {\affa}
\affiliation {\affb}
\author{\auttwo}
\affiliation {\affc}
\author{\autthree}
\affiliation {\affd}
%pr e uno

}{

\begin{center}
{\large {\bf \thetitle}}

\bigskip

\bigskip

%\catcode`@=11

{\large \noindent  \autone\footnote{jarah@impcas.ac.cn}{${}^{1,2}$}, \auttwo\footnote{abr84@psu.edu}{${}^{3}$} and \autthree\footnote{byzhang@henu.edu.cn}{${}^{4}$}}

%{\large \noindent  \autone{${}^{1,2}$} \footnote{jarah@impcas.ac.cn} and \auttwo{${}^{1,2}$} \footnote{guohengyuan@impcas.ac.cn}}

\vskip.7cm

1) \affa\\
2) \affb\\
3) \affc\\
4) \affd\\

\end{center}

}

\begin{abstract}
\noindent
We answer the question: If a vacuum sector Hamiltonian is regularized by an energy cutoff, how is the one-kink sector Hamiltonian regularized?  We find that it is not regularized by an energy cutoff, indeed normal modes of all energies are present in the kink Hamiltonian, but rather the decomposition of the field into normal mode operators yields coefficients which lie on a constrained surface that forces them to become small for energies above the cutoff.  This explains the old observation that an energy cutoff of the kink Hamiltonian leads to an incorrect one-loop kink mass.  To arrive at our conclusion, we impose that the regularized kink sector Hamiltonian is unitarily equivalent to the regularized vacuum sector Hamiltonian.  This condition implies that the two regularized Hamiltonians have the same spectrum and so guarantees that the kink Hamiltonian yields the correct kink mass.

\end{abstract}

% \vfill
%
% \end{titlepage}
\setcounter{footnote}{0}
\renewcommand{\thefootnote}{\arabic{footnote}}

\section{Introduction}

The conventional approach \cite{dhn2,rajaraman,physrept04,takyi}  to the computation of the mass of a quantum soliton is as follows.  One begins with a Hamiltonian which defines a theory and introduces a one-soliton sector Hamiltonian that describes the same physics in terms of the field expanded about the classical soliton solution.     Both Hamiltonians are then regularized.\footnote{{In the canonical transformation approach of \cite{gjs}, one applies the transformation to a perturbative sector Hamiltonian that includes counterterms.  The Hamiltonian is not regularized, however, because it is not expressed in terms of a regularized set of degrees of freedom -- Fourier modes up to a cut-off momentum, for example.}}  One defines the soliton mass to be the difference in the eigenvalues between a soliton ground state computed using the one-soliton sector Hamiltonian and the vacuum computed using the original vacuum Hamiltonian.  This mass depends on the regulators of both Hamiltonians.  Then the regulators must both be taken to infinity.  Unfortunately the mass found depends on the choice of how these two regulators are matched as they are taken to infinity \cite{rebhan}. 

Many prescriptions for this matching have been given in the literature, some leading to the right or the wrong answer.  In particular, a cut-off regularization of both Hamiltonians with a matching of the cut-off energies leads to the wrong answer \cite{rebhan}.  

Several proposals for resolving this problem have appeared in the literature.  The dependence on the regulator matching condition arises because of the sharp dependence of the energy on the regulator, which in turn is due to the quadratic ultraviolet divergence in the one-loop energies.  In Ref.~\cite{nastase} the authors noted that this problem could be removed by calculating not the energy itself, but rather its derivative with respect to an energy scale, as its divergence will be suppressed by one power.  In models simple enough for the constant of integration to be fixed using dimensional analysis, this allows the kink mass to become regulator independent and so one arrives at the right answer, even with the wrong regulator matching condition.  While this approach is sufficient for the calculation of the mass in many models of interest, it is not applicable to models with multiple scales and it cannot be used to calculate quantities with a steeper dependence on the energy scale.  

These shortcomings motivated Ref.~\cite{lit} to try to design a prescription for an energy cutoff that would reproduce the known one-loop kink mass.  This led them to an appealing physical principle, that normal modes sufficiently above the regulator scale should not be affected by the soliton.  Unfortunately this principle alone is not enough to fix the mass, as one must also choose how the density of states scales at high energies and the mass obtained depends on this choice. 

The origin of the above difficulties is clear.  In standard approaches, one transforms an unregularized defining Hamiltonian to the soliton sector and then regularizes, whereas one should apply the transformation directly to the regularized defining Hamiltonian.  For example, the original and accepted result for the one-loop correction to the soliton mass in $\phi^4$ theory \cite{dhn2} was obtained using mode number regularization.  Periodic boundary conditions for a box of size $L$ are applied to the relevant fluctuation operator in each sector, and the same number of modes is kept in determining the contribution to the energy.  This procedure is motivated by the lattice, but the full Hamiltonian was not defined on a lattice in \cite{dhn2}.  If it had been, and the map between sectors given within this defining theory, then there would be no ambiguity in determining how the perturbative and soliton sector regulators are related.  While $\phi^4$-theory has certainly been studied on a finite spatial lattice starting with \cite{Drell:1976bq}, where the authors were interested in the phase structure of the vacuum, a lattice version of the transformation from perturbative to soliton sectors has not been developed.

%\textcolor{teal}{One approach to resolve the ambiguity in the computation of quantum corrections to soliton masses, then, is to provide a finite-dimensional lattice version of the soliton-sector canonical transformation given in \cite{gjs,Tomboulis:1975gf}.  This is the approach we will follow in a forthcoming paper.  It is conceptually the most straightforward approach to resolving the old ambiguities, but it is technically nontrivial and might leave one wondering whether the field-theoretic answer depends on the regularization scheme.}

%\textcolor{teal}{One of the main messages of this paper is that the field-theoretic answer does not depend on the regularization scheme, as long as both the scheme and the regularized transformation are explicit so that regularization can be applied consistently in both sectors.  Then the computation is unambiguous and, by construction, each scheme will produce the same answer in the field-theoretic limit.  To illustrate this point, in this paper we apply a regularization scheme based on an energy cut-off rather than a mode number cut-off.  The novelty in both approaches, here and in the forthcoming work, is that the transformation to the soliton sector is applied to the fully regularized Hamiltonian.}

A satisfactory resolution to the above issue, then, is to provide a finite-dimensional lattice version of the soliton-sector canonical transformation given in \cite{gjs,Tomboulis:1975gf}.  This is the approach we will follow in a forthcoming paper.  It is conceptually straightforward but technically nontrivial, and one might ask if the same idea --transforming the regularized Hamiltonian to the soliton sector-- can be accomplished with a defining Hamiltonian regularized by an energy (or momentum) cutoff.  This paper answers in the affirmative.

Our work here follows in the footsteps of \cite{lit} in that we want to a obtain a consistent energy cut-off regularization and renormalization for kinks in 1+1 dimensions.  Like them, we recognize that the kink profile and the normal modes will depend on the regulator and in a consistent treatment these dependencies  should be considered, at least to be sure that they do not affect the answer to any given question.  However unlike them, we make no conjectures.  We derive our answer, as follows.

Consider a real scalar field $\phi(x,t)$.  Let $\phi(x,t)=f(x)$ be the classical kink solution.  The quantum theory will be treated in the Schrodinger picture, and so fields will be considered at a fixed $t$, and the $t$ argument will be dropped from the notation.  Now it is conventional to expand $\phi(x)$ about the classical solution as
\beq
\phi(x)=f(x)+\eta(x).
\eeq
In this case one could rewrite the Hamiltonian in terms of the quantum field $\eta(x)$ \cite{cahill76}.  

We will choose a different approach.  It is convenient to expand the field not about $f(x)$ but about a zero value of the field, {so that higher moments of the field correspond to interactions}.  This could be achieved using a passive transformation of the field $\phi\rightarrow \eta=\phi-f$.  Instead, following \cite{dhn2}, we will employ an active transformation of the Hamiltonian and momentum functionals which act on the field.  In particular we will transform the Hamiltonian as
\beq
H[\phi,\pi]\rightarrow H\p[\phi,\pi]=H[\phi+f,\pi]. \label{hpt}
\eeq
This definition of $H\p$ is sufficient for classical field theory, but in quantum field theory $H$ is regularized and we would like to define a regularized $H\p$.

The observation that underlies this paper is that (\ref{hpt}) is a unitary equivalence.  More specifically, we construct a unitary operator $\df$ which maps the vacuum sector to the kink sector \cite{hepp,sato}.  We define the regularized kink sector Hamiltonian $H^{\prime}$ by conjugating the defining, regularized vacuum-sector Hamiltonian $H$ with this operator $\df$
\beq
H^{\prime}=\df^\dag H \df. \label{sim}
\eeq
In this sense, the momentum cutoff $\Lambda$ in the kink sector is inherited from the vacuum sector.  Since the two Hamiltonians are similar,  they will have the same eigenvalues and their eigenvectors, which differ by the operator $\df$, represent the same states.  The kink mass is the difference between the eigenvalues of the eigenstates corresponding to the vacuum and the kink ground state.  These two eigenstates may be identified in either Hamiltonian, as they have the same eigenvalues and states, however in perturbation theory we may only find the vacuum as an eigenvector $|\Omega\rangle$ of the vacuum Hamiltonian and the kink ground state $|K\rangle$ as an eigenvector $\vac$ of the kink Hamiltonian.  In other words, once we have used perturbation theory to find the kink ground state as an eigenstate $\vac$ of $H^{\prime}$ then the corresponding eigenstate of the defining Hamiltonian $H$ is 
\beq
|K\rangle=\df \vac. \label{co}
\eeq

Note that this approach is not sensitive to the exact choice of $\df$.  Any $\df$ which allows one to find the desired eigenvectors of $H^{\prime}$ is sufficient, since they all have the same eigenvalues and the eigenvectors are easily mapped between the various eigenspaces using $\df$.

% one only requires that it takes the vacuum $|\Omega\rangle$ close enough to the kink ground state $|K\rangle$ that the mismatch is an operator $\co$ 
%\beq
%|K\rangle=\df \co|\Omega\rangle \label{co}
%\eeq
%that may be calculated perturbatively.

%The key realization is that this same construction may be applied to the regularized Hamiltonian $H^{\L}$.  The kink will be slightly deformed by the regularization and so $\df$ will be deformed, but it will still be unitary and so
%\beq
%H^{\prime\L}=\df^\dag H^{\L}\df
%\eeq
%will still have the same spectrum as  $H^{\L}$ and the eigenvalue difference will be the regularized kink mass.  

The regulator $\Lambda$ can then be taken to infinity unambiguously, using the renormalization conditions, to arrive at the renormalized kink mass.  This limit is unambiguous as there is only one regulator $\Lambda$ which needs to be taken to infinity, not two.  The similarity transformation (\ref{sim}) guarantees the correct regulator for the kink Hamiltonian.  Previous approaches to an energy cutoff in these models failed\footnote{In Ref.~\cite{rebhan} it was shown that they yield the wrong one-loop kink mass.} because they regularized the kink sector Hamiltonian by imposing an energy cutoff in the spectrum of normal modes, but we will see that $H^{\prime}$ includes all normal modes.

In this work we only include an ultraviolet (UV) regulator since the computations considered will not require infrared (IR) regularization.  In forthcoming work we will present a soliton-sector transformation for a finite-dimensional lattice regularization that uses both.  

Infrared (IR) regularization via compactification or the inclusion of antisolitons is common in the literature.  However, such IR regularization is necessarily different for the vacuum and soliton sectors, which makes it difficult to treat nonperturbative effects which mix distinct sectors.  Mixing may seem unimportant at weak coupling, but at strong coupling it is responsible for the symmetry restoring phase transition in the $\phi^4$ theory.  Also the dual Thirring description for the Sine-Gordon model includes a mixing of two-fermion and zero-fermion states.

We begin in Sec.~\ref{vacsez} by defining the regularized theory and the vacuum sector.  Next in Sec.~\ref{mapsez} we define the similarity transform which takes the defining Hamiltonian to the kink sector Hamiltonian.  Finally in Sec.~\ref{kinksez} we find the regularized kink sector Hamiltonian and calculate its one-loop ground state and mass.  When the regulator is taken to infinity, we reproduce the known mass formula.  Our notation is summarized in Table~\ref{notab}.

\begin{table}
\begin{tabular}{|l|l|}
\hline
Operator&Description\\
\hline
$\phi^\L(x),\ \pi^\L(x)$&The real scalar field and its conjugate momentum\\
$A^\ddag_p,\ A_p$&Creation and annihilation operators in plane wave basis\\
$B^\ddag_k,\ B_k$&Creation and annihilation operators in normal mode basis\\
$B^\ddag_{BO},\ B_{BO}$&Creation/annihilation operators of odd shape modes\\
$\phi_B,\ \pi_B$&Zero mode of $\phi(x)$ and $\pi(x)$ in normal mode basis\\
$::_a$&Normal ordering with respect to $A$ operators respectively\\
%$S[]$&Symmetrization with respect to momenta\\
\hline
%Indices&Description\\
%\hline
%$m$&Contractions\\
%$i$&shape modes\\
%$I$&Bound states including both shape modes and the zero mode\\
%$M$&Normal mode type: zero mode, shape or continuum mode\\
\hline
Hamiltonian&Description\\
\hline
$H$&The original Hamiltonian\\
$H^\prime$&$H$ with $\phi^\L(x)$ shifted by regularized kink solution $f^\L(x)$\\
$H\p_n$&The $\phi^{\L n}$ term in $H^\prime$\\
\hline
Symbol&Description\\
\hline
$\Lambda$&Momentum cutoff\\
$f_0(x)$&The unregularized classical kink solution\\
$f_1(x)$&First correction to the unregularized classical kink solution\\
$f^\L(x)$&Regularized classical kink solution\\
$\df$&Operator that translates $\phi^\L(x)$ by the regularized classical kink solution\\
$g_B(x)$&The kink linearized translation mode\\
%$g_{BE,i}(x),\ g_{BO,i}(x)$&The $i$th even/odd shape mode\\
$g_k(x)$&Normal modes including discrete modes\\
%$\gamma_i^{mn}$&Coefficient of $\phi_0^m B^{\dag n}\vac_0$ in order $i$ ground state\\
%$\Gamma_i^{mn}$&Coefficient of $\phi_0^m B^{\dag n}\vac_0$ in order $i$ Schrodinger Equation\\
%$V_{ijk}$&Derivative of the potential contracted with various functions\\
%$Y_{ijk}$&$V_{ijk}$ divided by a sum of frequencies\\
%$\I(x)$&Contraction factor from Wick's theorem\\
$p$&Momentum\\
$k$&Normal mode label\\
$\omega_k,\ \omega_p$&The frequency corresponding to $k$ or $p$\\
%$\Omega_i$&Sum of frequencies $\ok{}$\\
$\tilde{g}$&Inverse Fourier transform of $g$\\
%$Q_n$&$n$-loop correction to kink energy\\
%$\hat{g}$&Fourier transform of $\tilde{g}/\omega$ \\
%$I_M(x)$&Contraction arising from type $M$ normal mode\\
%$N_k,\ N^M_k$&Plane wave normal ordered product of $k$ $A^\ddag+a$ or $A^\ddag_M+a_M$ factors\\
%$B^M_n$&Normal mode normal ordered product of $n$\ $B^\ddag\pm b$ factors\\
%$\alpha_{nm},\ a_{nm}$&Dimensionful/less coefficients for $n$ field products with $m$ contractions\\
\hline
State&Description\\
\hline
$|K\rangle, \ |\Omega\rangle$&Kink and vacuum sector ground states\\
%$$&True ground state\\
%$\co_n|\Omega\rangle$&Translation of $|K\rangle$ by $\df^{-1}$  at order $n$ \\
\hline

\end{tabular}
\caption{Summary of Notation}\label{notab}
\end{table}

\section{The Vacuum Sector} \label{vacsez}

\subsection{Classical Theory}

Let us consider a classical theory of a real scalar field $\tilde{\phi}^\L(x)$ and its conjugate $\pi^\L(x)$ in 1+1 dimensions,
\beq
H=\int dx \mh\hsp
\mh=\frac{\pi^{\L 2}(x)+\partial_x \tilde{\phi}^\L (x)\partial_x \tilde{\phi}^\L (x)}{2}+\frac{\tilde{V}[g\tilde{\phi}^\L]}{g^2}+\frac{\delta m^2}{2}\tilde{\phi}^{\L 2}+\tilde{\gamma}
\eeq
where we have included counterterms $\delta m^2$ and $\tilde{\gamma}$ and a potential which in the case of the $\phi^4$ double well theory is  
\beq
\tilde{V}[g\tilde{\phi}^\L]=\frac{1}{4}\left(g^2\tilde{\phi}^{\L 2}-g^2v^2\right)^2.
\eeq
The potential is always written as a function of $g\phi$ because this combination is dimensionless.  The notation $\L$ emphasizes that the quantity in question depends on the regulator $\Lambda$.   In this note we will focus on the $\phi^4$ theory, where renormalization conditions can be chosen such that the coupling is not renormalized and the only required counterterms are those proportional to $\delta m^2$ and $\tilde{\gamma}$ \cite{rebhan}.  However the strategy can be readily generalized to other potentials by adding counterterms for the additional interactions and corresponding renormalization conditions.

We will be interested in cases in which $V[g\phi]$ has multiple minima, so that there is a kink solution in the classical theory.  To simplify the discussion below, we will shift the field by the location of one of these minima, $-v$ by defining
\beq
\phi^\L(x)=\tilde{\phi}^\L(x)+v\hsp
V[g\phi^\L]=\tilde{V}[g(\phi^\L-v)]\hsp
\gamma=\tilde{\gamma}+\frac{\delta m^2}{2}v^2 \label{shift}
\eeq
leading to
\beq
\mh=\frac{\pi^{\L 2}(x)+\partial_x {\phi}^\L (x)\partial_x {\phi}^\L (x)}{2}+\frac{{V}[g{\phi}^\L]}{g^2}+\frac{\delta m^2}{2}{\phi}^{\L 2}-v\delta m^2\phi^\L(x)+{\gamma}. \label{h}
\eeq
In the case of the $\phi^4$ double well one obtains
\beq
{V}[g{\phi}^\L]=\frac{g^2\phi^{\L 2}}{4}\left(g\phi^{\L}-2gv\right)^2.
\eeq

\subsection{Quantum Theory} \label{qsez}

As the action has the dimensions of $\hbar$, the operator $\phi^\L(x)$ has dimensions $O(\hbar^{1/2})$ while $g$ has dimensions  $O(\hbar^{-1/2})$.  Therefore $v$ also has dimensions $O(\hbar^{1/2})$.  This means that in the quantum theory, they will appear in the dimensionless combinations $g\hbar^{1/2}$ and $v\hbar^{-1/2}$.  Thus in the semiclassical expansions below, each power of $v$ will be treated as one power of $1/g$.

We will not directly quantize $\phi^\L(x)$ and $\pi^\L(x)$.  Instead we will decompose
\beq
\phi^\L(x)=\lpin{p} \phi_p e^{-ipx}\hsp
\pi^\L(x)=\lpin{p} \pi_p e^{-ipx} \label{phip}
\eeq
and we will quantize $\phi_p$ and $\pi_p$ by imposing
\beq
[\phi_p,\pi_q]=2\pi i \delta(p+q). \label{cr}
\eeq
This note will be entirely in the Schrodinger picture, and so (\ref{cr}) defines Schrodinger operators $\phi^\L(x)$ and $\pi^\L(x)$.  They are not local fields; indeed they satisfy
\beq \label{nonlocalccr}
\left[\phi^\L(x),\pi^\L(y)\right]=i\lpin{p}e^{-ip(x-y)}=i\alpha(x-y)\hsp
\alpha(x)=\frac{\sin (\Lambda x)}{\pi x}.
\eeq
The nonlocality of the commutator arises from the fact that the integration in the momentum space is cut off by $\Lambda$.  Note that locality is restored as the cutoff is removed because
\beq
\stackrel{\rm{lim}}{{}_{\Lambda\rightarrow\infty}}\alpha(x)=\delta(x). \label{alim}
\eeq

Interestingly, \eqref{nonlocalccr} are the same canonical commutation relations found in Pearson's thesis \cite{pearson} (see also \cite{dwy}) for an infinite lattice regularization with lattice spacing $2\pi/\Lambda$.  In this context, $\phi^{(\Lambda)}, \pi^{(\Lambda)}$ are interpolating fields for underlying lattice degrees of freedom that do satisfy canonical commutation relations.  Using the interpolating fields to define the regularized Hamiltonian, as we have done here, implies that our theory is equivalent to the infinite lattice theory of Pearson.  We do not utilize the lattice language in this paper since it is not advantageous for the computations we perform.  We will utilize it, however, in forthcoming work when we construct a finite-dimensional version of the soliton-sector canonical transformation.

Our problem is now well-defined.  One can insert (\ref{phip}) into (\ref{h}) to obtain the Hamiltonian as a function of the operators $\phi_p$ and $\pi_p$.  The vacuum and the kink ground state will be eigenstates of this Hamiltonian, and the difference between their eigenvalues is the kink mass.  The Schrodinger picture is sufficient for determining the spectra and so we need never introduce time, and the nonlocality of our operators will not impede our narrow task.

To do this concretely, we will choose the renormalization condition 
\begin{equation} \label{rc}
H |\Omega \rangle = 0~
\end{equation}
and will expand
\beq
|\Omega \rangle = \sum_{i=0}^{\infty} |\Omega \rangle_i\hsp|\Omega \rangle_i = O(g^{2i})\hsp
\gamma=\sum_{i=0}^{\infty} \gamma_i\hsp
\gamma_i = O(g^{2i}). \label{gam}
\eeq
%
%\beq
%H|\Omega\rangle=E|\Omega\rangle\hsp
%E=\sum_{i=0}^{\infty} E_i\hsp 
%E_i=O\left(g^{2i}\right). \label{rc}
%\eeq
Let us begin with the case $i=0$.  We are interested in terms in $\mh$ with no powers of $g$.   Let us assume for the moment that $\delta m^2$ contains only terms of order at least $O(g^2)$.  We will see shortly that this choice is consistent.  Then we are left with
\bea
H_{0}&=&\int dx \mh_{0}\label{h0}, \\
\mh_{0}&=&\frac{\pi^{\L 2}(x)+\partial_x {\phi}^\L (x)\partial_x {\phi}^\L (x)+m^2\phi^{\L 2}(x)}{2}+{\gamma}_0\hsp m=\sqrt{{V}^{\prime\prime}[0]}. \nonumber
\eea
For the $\phi^4$ double well, $m=\sqrt{2}vg$.  Define the linear combinations
\beq
A^\ddag_p=\frac{\phi_p}{2}-i\frac{\pi_p}{2\omega_p}\hsp
A_{-p}=\omega_p\phi_p+i\pi_p\hsp\omega_p=\sqrt{m^2+p^2} \label{adef}
\eeq
where the Hermitian conjugate of $A_p$ is $2\omega_p A^\ddag_p$.   The canonical commutation relations (\ref{cr}) imply that these each satisfy a Heisenberg algebra
\beq
[A_p,A^\ddag_q]=2\pi\delta(p-q).
\eeq
To impose (\ref{rc}) at $O(g^0)$, it suffices to consider $ |\Omega\rangle_0$ which is the free vacuum
\beq
A_p |\Omega\rangle_0=0.
\eeq

The definitions (\ref{adef}) are easily inverted
\beq
\phi_p=A^\ddag_p+\frac{A_{-p}}{2\omega_p}\hsp
\pi_p=i\left(\omega_pA^\ddag_p-\frac{A_{-p}}{2}\right).
\eeq
Substituting this into (\ref{h0}) one finds
\beq
0=H_{0}|\Omega\rangle_0=\int dx \left(\gamma_0+\lpin{p}\frac{\omega_p}{2}\right)|\Omega\rangle_0.
\eeq
and so the leading order counterterm is
\beq
\gamma_0=-\lpin{p}\frac{\omega_p}{2}=-\frac{1}{4\pi}\left[\Lambda \omega_\Lambda+m^2 {\rm{ln}}\left(\frac{\Lambda+\omega_\Lambda}{m}
\right)\right] 
\eeq
where we have integrated by parts.  We have now satisfied the renormalization condition (\ref{rc}) at order $O(g^0)$. 

%\blu{To include Andy's proposed comment about higher order corrections, and the fact that higher order tadpoles have loops and so the condition can't be read directly off of the Hamiltonian without evaluating the loops, I have reworked the following paragraph.  Please check to see if you agree.} 

To fix $\delta m^2$ we will impose another renormalization condition, that tadpoles vanish.  Recall that $\delta m^2$ is a series in $g$ beginning at order $O(g^2)$. For the computation of these leading order terms, the no-tadpole condition is equivalent to imposing that $H|\Omega\rangle_0$ contains no terms of the form $A^\ddag|\Omega\rangle_0$ at $O(g)$.  The relevant $O(g)$ terms in the Hamiltonian are
\beq
H_{1}=\int dx \mh_{1}\hsp
\mh_{1}=g\frac{V^{(3)}[0]\phi^{\L 3}(x)}{6}-v\delta m^2\phi^\L(x) \label{h1}
\eeq
where $V^{(n)}[0]$ is the $n$th derivative of $V$ and we recall that $v$ is of order $O(1/g)$.% and we will consider the leading $O(g^2)$ part of $\delta m^2$, \textcolor{red}{though in principle $\delta m^2$ also has a series of corrections in $g$.}   

Using the terms with a single $A^\ddag$
\beq
\int dx \phi^\L(x)|\Omega\rangle_0=A^\ddag_0|\Omega\rangle_0\hsp
\int dx \phi^{\L 3}(x)|\Omega\rangle_0\supset 3\lpin{p}\frac{1}{2\omega_p}A^\ddag_0|\Omega\rangle_0
\eeq
one obtains
\beq
\delta m^2=\frac{g}{2v}V^{(3)}[0]\lpin{p}\frac{1}{2\omega_p}=\frac{g}{4\pi v}V^{(3)}[0]{\rm{ln}}\left(\frac{\Lambda+\omega_\Lambda}{m}\right).
\eeq
In the case of the $\phi^4$ double well this is
\beq
\delta m^2=-\frac{3g^2}{2\pi}{\rm{ln}}\left(\frac{\Lambda+\omega_\Lambda}{m}\right).
\eeq

\subsection{The Relation to Normal Ordering}

One may define a normal ordering on the operators $A^\ddag$ and $A$ by placing all of the former on the left.  We will denote this normal ordering $:O:_a$ for any operator $O$.  It is easily evaluated on the operators appearing in the Hamiltonian at the orders considered above
%{\bf{NO .. considera la dipendenza da x}}
\bea
\pi^{\L 2}(x)&=&:\pi^{\L 2}(x):_a+\lpin{p}\frac{\omega_p}{2},\\
\phi^{\L 2}(x)&=&:\phi^{\L 2}(x):_a+\lpin{p}\frac{1}{2\omega_p}\nonumber,\\
\left(\partial_x\phi^{\L}(x)\right)^2&=&:\left(\partial_x\phi^{\L}(x)\right)^2:_a+\lpin{p}\frac{p^2}{2\omega_p}\nonumber,\\
\phi^{\L 3}(x)&=&:\phi^{\L 3}(x):_a+3\phi^\L(x)\lpin{p}\frac{1}{2\omega_p}\nonumber.
\eea
Inserting these into (\ref{h0}) and (\ref{h1}) one finds
\bea
\mh_{0}&=&\frac{:\pi^{\L 2}(x)+\partial_x {\phi}^\L (x)\partial_x {\phi}^\L (x)+m^2\phi^{\L 2}(x):_a}{2} \nonumber\\
\mh_{1}&=&g\frac{V^{(3)}[0]:\phi^{\L 3}(x):_a}{6}.
\eea

We see that order by order, properly chosen renormalization conditions are equivalent to a Hamiltonian that is normal ordered from the beginning.  More specifically, the required renormalization condition sets to zero all diagrams with a loop involving a single vertex. In the case of two-dimensional scalar field theories, normal ordering is sufficient to remove all divergences (see for example Ref.~\cite{colemansg}).  Thus it would be possible to take the limit $\Lambda\rightarrow\infty$ now.  This would reduce the problem to that solved already in Ref.~\cite{mekink,memassa} at one loop and \cite{dueloop} at two loops where it was shown that one arrives at the correct kink mass.  However, we will not follow this approach as it will not generalize to theories with fermions and theories in higher dimensions, which motivate this line of research.

In the case of the $\phi^4$ theory, with the renormalization conditions of Subsec.~\ref{qsez},  the total Hamiltonian density is then
\begin{equation}
\mh=\frac{:\pi^{\L 2}(x):_a+:\partial_x {\phi}^\L (x)\partial_x {\phi}^\L (x):_a}{2}+\frac{:{V}[g{\phi}^\L]:_a}{g^2} + \tilde{\gamma}_1+\sum_{i=2}^{\infty}\gamma_{i}. \label{rh}
\end{equation}
Here we have shifted the subleading counterterm to absorb the contribution from Wick contractions appearing in the normal ordering of the $\phi^4$ interaction %\textcolor{red}{This looks good, but is it not true that $\tilde{\gamma}_1 = 0$ by our renormalization condition?  Or are you saying you found it needed to be nontrivial in your 2-loop computation?} \blu{Yes, if I remember correctly there is a 2-loop correction to the vacuum energy density in the $\phi^4$ theory (it starts at three loops in Sine-Gordon).  It comes from the diagram with two vertices, connected by 3 lines.  So $\tilde{\gamma}_1$ needs to be chosen to cancel this, to satisfy our renormalization condition. I have now added a reference just below the equation}
\beq
\tilde{\gamma}_1=\gamma_1-\frac{3g^2}{4}\left(\lpin{p}\frac{1}{2\omega_p}\right)^2.
\eeq
In Refs.~\cite{phi42loop,phi42loop2}, the calculation of $\tilde{\gamma}_1$ was reviewed and it played an essential role in the elimination of IR divergences in the calculation of the two-loop kink mass.  The shifted counterterm and all successive terms are finite at $\Lambda\rightarrow\infty$, with the first few given in Eq.~(47) of Ref.~\cite{serone}.  In the case of a general potential, normal ordering again leads to such infinite shifts and finite remainders.

\section{Defining the Kink Sector} \label{mapsez}

\subsection{The Similarity Transformation}

Let $f(x)$ be a real-valued function.  Define the displacement operator
\beq
\df={\rm{exp}}\left(-i\int dx f(x)\pi^\L(x)\right) \label{df}
\eeq
and
\beq
\tf(p)=\int dx f(x) e^{ipx}.
\eeq
This integral is often divergent.  Our prescription for defining $\tf(p)$ in this case is described in \ref{appOne}.

Exponentiating the commutators
\beq
\left[\int dx f(x) \pi^\L(x),A_q\right]=i\pin{p} \omega_p\tf(-p)\left[A^\ddag_p,A_q\right]=-i \omega_q\tf(-q)
\eeq
and
\beq
\left[\int dx f(x) \pi^\L(x),A^\ddag_q\right]=-i\pin{p}\frac{1}{2}\tf(-p)\left[A_{-p},A^\ddag_q\right]=-\frac{i}{2}\tf(q)
\eeq
one finds
\beq
\left[\df,A_p\right]=-\omega_p\tf(-p) \df\hsp
\left[\df,A^\ddag_p\right]=-\frac{\tf(p)}{2} \df.
\eeq
Therefore
\beq
[\df,\phi^\L(x)]=\lpin{p}\left([\df,A^\ddag_p]+\frac{[\df,A_{-p}]}{2\omega_p}\right)e^{-ipx}
=-f^\L(x)\df
\eeq
where
\beq
f^\L(x)=\lpin{p}\tf(p)e^{-ipx} \label{fLsimpledef}
\eeq
satisfies $f^\L(x)=f(x)$ if $\tf(q)=0$ for all $q$ with $|q|>\Lambda$.  Note that $\df$ commutes with $\pi^\L$ as $[\pi_p,\pi_q]=0$.

Finally we are ready to introduce the similarity transformation at the heart of our construction.  The kink sector Hamiltonian $H\p$ is defined by
\beq 
\df^\dag H[\phi^\L(x),\pi^\L(x)]\df=H\p[\phi^\L(x),\pi^\L(x)]=H[\phi^\L(x)+f^\L(x),\pi^\L(x)] \label{hp}
\eeq
for a suitable choice of $f(x)$, which will be made momentarily.  In the rest of this note we will be concerned with $H\p$. As it is similar to $H$, it has the same spectrum.  In particular if
\beq
H|K\rangle=E_K|K\rangle
\eeq
then, using (\ref{co}),
\beq
H\p\vac=E_K\vac
\eeq
and so $H\p$ may be used to compute the energy of any state, and in particular the energy $E_K$ of the kink ground state.

\subsection{The Kink-Sector Tadpole}

Let us consider $f^\L(x)$ to be of order $O(1/g)$.  Then the kink sector Hamiltonian $H\p$ can be expanded order by order in the coupling $g$
\beq
H\p=\sum_{i=0}^\infty H\p_i\hsp
H\p_i=\int dx \mh\p_i
\eeq
where each $H\p_i$ and $\mh\p_i$ is of order $O(g^{i-2})$.  At leading order one finds the classical energy corresponding to the field configuration $\phi^\L(x)=f^\L(x)$ 
\beq
\mh\p_0=\frac{\left(\partial_x f^\L(x)\right)^2}{2}+ \frac{V[g f^{\L }(x)]}{g^2}.
\eeq

The kink-sector tadpole appears at the next order:
\beq
H\p_1=\int dx \phi^\L(x)\left[-\partial^2_xf^{\L}(x)+V\p[gf^\L(x)]/g\right]. \label{tad}
\eeq
We will define $f^\L$ by imposing that the expression in brackets vanishes at momenta below the cutoff\footnote{Note that, although the higher modes of $f^{\L}$ vanish, those of the $V\p$ term do not vanish.  However they do not contribute to the tadpole in Eq.~(\ref{tad}) as the higher modes of $\phi^\L$ vanish.}
\beq
\int dx \left[-\partial^2_xf^{\L}(x)+V\p[gf^\L(x)]/g\right]e^{ipx}=0\hsp |p|\leq \Lambda \label{fdef}
\eeq
which will automatically eliminate the kink sector tadpole as $\phi^\Lambda(x)$ satisfies the opposite condition
\beq
\int dx \phi^\L(x) e^{ipx}=0\hsp |p|> \Lambda.
\eeq
In the case $\Lambda\rightarrow \infty$ this condition corresponds to the classical equations of motion for a time-independent configuration in the unregularized theory.  Thus $f^{(\infty)}(x)$ will be equal to the classical kink solution $f_0(x)$ and $\mh\p_0$ will reduce to its classical mass density.

The definition (\ref{fdef}) can be solved as an expansion in large $\Lambda$ about the classical solution $f_0(x)$.  To illustrate this procedure, in \ref{appOne} we find the leading correction in the case of the $\phi^4$ double well and show that it is exponentially small in $\Lambda/m$.

\section{The Kink Sector} \label{kinksez}

\subsection{The Setup}

In this section we will restrict our attention to the $\phi^4$ double well theory.  As noted above, an analogous treatment of models with other interactions generically requires additional counterterms multiplying various powers of $\phi^\L$.  Inserting (\ref{rh}) into (\ref{hp}) one finds that the $g$-independent terms are
\beq
H\p_2=\int dx \frac{:\pi^{\L 2}(x):_a+:\partial_x {\phi}^\L (x)\partial_x {\phi}^\L (x):_a+V\pp[g f^\L(x)]:\phi^{\L 2}(x):_a}{2}. \label{h2}
\eeq
These are the only terms that contribute at one loop, as they are suppressed by a factor of $g^2$ with respect to the leading terms in $H\p_0$ and we have chosen $f^\L(x)$ so that $H\p_1$ vanishes.  The term $g f^{(\Lambda)}$ is $g$-independent and becomes $g f_0$ as $\Lambda \to \infty$. However, we argued in \ref{appOne} that the corrections $f^\L-f_0$ are exponentially small in $m/\Lambda$.  Since perturbative computations only produce power-law divergences in $\Lambda$, such exponentially small corrections do not contribute to any perturbative quantities.  Therefore in the following we will replace $V''[g f^{(\Lambda)}(x)]$ with $V''[g f_0(x)]$.

For completeness we note 
\bea
H\p_3&=&\int dx\frac{gV^{(3)}[g f^\L(x)]:\phi^{\L 2}(x):_a}{6}\\
H\p_4&=&\int dx\left[\frac{g^2V^{(4)}[g f^\L(x)]:\phi^{\L 2}(x):_a}{24}+\tilde{\gamma}_1\right].\nonumber
\eea

In the rest of this note we will use (\ref{h2}) to study the one-loop kink spectrum.  This equation describes a theory which is free, as the Hamiltonian is quadratic in the field.  However the operators $A$ and $A^\ddag$ do not diagonalize the Hamiltonian as a result of the $x$-dependent mass term.  Our strategy will thus be to diagonalize $H$ using a Bogoliubov transformation from the basis of operators that create plane waves to a basis of operators that create normal modes of the regularized kink.

First let us find these normal modes.  Inserting the Ansatz\footnote{This constant frequency Ansatz is inconsistent with the restriction that the higher Fourier modes of $\phi^\L$ vanish.  However, when we construct the quantum field $\phi^\L$ below, we will consider only those linear combinations which satisfy the restriction.  The key observation will be that these combinations do not have constant frequency, and so our regularized kink Hamiltonian is not obtained by cutting off frequencies above any sharp threshold.}
\beq
\phi^\L(x,t)=e^{i\omega_k t}g_k(x)
\eeq
into the classical equations of motion for (\ref{hp}) one finds
\beq
V\pp[gf_0(x)]g_k(x)=(\omega_k^2+\partial_x^2)g_k(x). \label{fluceqn}
\eeq
The index $k$ will include a continuous spectrum with energy
\beq
\omega_k=\sqrt{m^2+k^2}
\eeq
as well as discrete modes.  In the case of the $\phi^4$ modes there are two discrete solutions, the zero mode $g_B(x)$ corresponding to the translation symmetry with $\omega_B=0$ and also an odd shape $g_{BO}(x)$ with $\omega_{BO}=m\sqrt{3}/2$.

As $V$ is real we may choose
\beq
g^*_k(x)=g_k(-x)
\eeq
and so $g_B(x)$ is real, $g_{BO}(x)$ is imaginary and in the case of continuum modes
\beq
g^*_k(x)=g_{-k}(x).
\eeq
We normalize the solutions such that
\beq
\int dx |g_B(x)|^2=\int dx |g_{BO}(x)|^2=1
\eeq
and in the case of continuum modes
\beq
\int dx g_{k_1}(x) g_{k_2}(x)=2\pi\delta(k_1+k_2).
\eeq
These satisfy the completeness relation
\beq
g_B(x)g^*_B(y)+g_{BO}(x)g^*_{BO}(y)+\pin{k}g_k(x)g_{-k}(y)=\delta(x-y). \label{comp}
\eeq
From here on we will adopt the shorthand that $\pin{k}$ implicitly includes a sum over the discrete modes $g_B$ and $g_{BO}$.  We also introduce the inverse Fourier transform
\beq
\tilde{g}_k(p)=\int dx g_k(x) e^{ipx}.
\eeq

The completeness relation (\ref{comp}) implies that the functions $g_k(x)$ are a basis of all functions so we may decompose
\beq
\phi^\L(x)=\lpin{p} \phi_p e^{-ipx}=\pin{k}\phi^\L_k g_k(x)
\hsp
\pi^\L(x)=\lpin{p} \pi_p e^{-ipx}=\pin{k}\pi^\L_k g_k(x). \label{dec}
\eeq
Recall that the $k$ integral implicitly sums over bound states, and so for example we have also introduced $\phi_B$ and $\pi_B$.  % which we will denote $\phi_0$ and $\pi_0$ as they represent zero modes.
Note that the $k$ integrals are never cut off.

As the $p$ integral is bounded, the field satisfies the constraint
\beq
\int dx \phi^\L(x) e^{iqx}=\int dx \pi^\L(x) e^{iqx}=0\hsp |q|>\Lambda. \label{cons}
\eeq
The $k$ integral is not bounded, indeed including the discrete sums it runs over a complete basis.  Therefore (\ref{cons}) implies a constraint on the coefficients
\beq
0= \pin{k}\phi^\L_k \tilde{g}_k(q)= \pin{k}\pi^\L_k \tilde{g}_k(q)\hsp |q|>\Lambda.  \label{ccons}
\eeq

Here we see the difference between our approach and the traditional mode matching \cite{dhn2} or energy cut-off \cite{rebhan} regularization schemes.  {\it{In the traditional approach, one limits the $k$ integration to include either the same number of modes as in the $p$ integration or else an integral out to the same energy.  Here instead we integrate over all $k$ but with a constraint on the coefficients $\phi^\L_k$ which leads these coefficients to be small\footnote{The $\phi^\L_k$ are operators.  They are small at large $|k|$ in the sense that they are superpositions of $\phi_p$ with small coefficients.} at $k>\Lambda$ but nonzero}}.  We claim that this is the correct way to regularize the kink Hamiltonian with a UV energy cutoff because in this approach the regularized kink Hamiltonian is related by a similarity transformation to the vacuum Hamiltonian, which defines the theory. In particular they have the same spectrum, and thus the kink mass, which is the difference between two eigenvalues, can be calculated using one eigenvalue from each Hamiltonian.

The decompositions lead to the Bogoliubov transformations
\beq
\phi_p= \pin{k}\phi^\L_k \tilde{g}_k(p)\hsp
\pi_p=\pin{k}\pi^\L_k \tilde{g}_k(p).
\eeq
Integrating (\ref{dec}) over $x$ with weight\footnote{If $k$ is a discrete mode, $g_{-k}=(-1)^Pg_k$ where $P$ is the parity of the discrete mode $k$.} $g_{-k}(x)$ one finds the inverse transformations
\beq
\phi^\L_k=\lpin{p}\phi_p\tilde{g}_{-k}(-p)\hsp
\pi^\L_k=\lpin{p}\pi_p\tilde{g}_{-k}(-p). %controlla
\eeq
%It may seem strange that the $\phi^\L_k$, which have an infinite measure, can be determined  in terms of the $\phi_p$ which have measure of order $\Lambda$.  This is possible because the $\phi^\L_k$ always are valued on the constrained surface (\ref{cons}) which has the same measure as the $\phi_p$.  \textcolor{red}{I find this last comment more confusing than helpful.  I don't think it's needed.}
%The Fourier transform of the completeness relation (\ref{comp}) 
%\beq
%\pin{k}\tilde{g}_k(p)\tilde{g}_{-k}(q)=2\pi\delta(p+q)
%\eeq
%leads to the inverse
%\beq
%\phi_k=

\subsection{Finding the One-Loop Hamiltonian}

$H\p_2$ consists of three terms
\beq
H\p_2=A+B+C.
\eeq
First consider the potential term
\bea
A&=&\int dx \frac{V\pp[g f^\L(x)]:\phi^{\L 2}(x):_a}{2}\\
&=&\int dx\int dy \frac{V\pp[g f^\L(x)]\delta(x-y):\phi^{\L}(x)\phi^\L(y):_a}{2}\nonumber\\
&=&\int dx\int dy\pin{k} \frac{V\pp[g f^\L(x)]g_k(x)g^*_k(y):\phi^{\L}(x)\phi^\L(y):_a}{2}\nonumber\\
&=&\int dx\int dy\pin{k} \frac{\left[(\omega_{k}^2+\partial_x^2)g_k(x)\right]g^*_k(y):\phi^{\L}(x)\phi^\L(y):_a}{2}.\nonumber
\eea
The derivative term cancels
\bea
B&=&-\int dx \frac{:{\phi}^\L (x)\partial^2_x {\phi}^\L (x):_a}{2}\\\notag
&=&-\int dx\int dy \pin{k} \frac{g_k(x)g^*_{k}(y):{\phi}^\L (y)\partial^2_x {\phi}^\L (x):_a}{2}\nonumber
\eea
after integrating $x$ by parts twice, leaving
\bea
A+B&=&\int dx\int dy\pin{k} \frac{\omega_{k}^2 g_k(x)g^*_k(y):\phi^{\L}(x)\phi^\L(y):_a}{2}\\
&=&\pin{k}\lpin{p_1}\lpin{p_2} \frac{\omega_{k}^2 \tilde{g}_k(-p_1)\tilde{g}^*_k(p_2):\phi_{p_1}\phi_{p_2}:_a}{2}=D+E\nonumber
\eea
where
\bea
D&=&\pin{k}\lpin{p_1}\lpin{p_2} \frac{\omega_{k}^2 \tilde{g}_k(-p_1)\tilde{g}^*_k(p_2)\phi_{p_1}\phi_{p_2}}{2}\label{deq}\\
&=&\pin{k}\omega_k^2\frac{\phi^\L_{k}\phi^\L_{-k}}{2}.\nonumber
\eea
Here in the case of a discrete mode $k$ with parity $P$, $\phi_{-k}=(-1)^P\phi_k$.  The contraction term is
\bea
E&=&-\pin{k}\lpin{p_1}\lpin{p_2} \frac{\omega_{k}^2 \tilde{g}_k(-p_1)\tilde{g}^*_k(p_2)}{2}\frac{2\pi\delta(p_1+p_2)}{2\omega_{p_1}}\\
&=&-\pin{k}\lpin{p}\tilde{g}_k(p)\tilde{g}^*_k(p) \frac{\omega_{k}^2 }{4\omega_p}.\nonumber
\eea

The last term in $H\p_2$ is
\bea
C&=&\int dx \frac{:\pi^{\L 2}(x):_a}{2}
=\int dx\int dy \delta(x-y) \frac{:\pi^{\L}(x)\pi^\L(y):_a}{2}\\
&=&\int dx\int dy\pin{k}\frac{g_k(x)g^*_k(y):\pi^{\L}(x)\pi^\L(y):_a}{2}\nonumber\\
&=&\int dx\int dy\pin{k}\lpin{p_1}\lpin{p_2}
\frac{g_k(x)g^*_k(y)e^{-ip_1x-ip_2y}:\pi_{p_1}\pi_{p_2}:_a}{2}\nonumber\\
&=&\pin{k}\lpin{p_1}\lpin{p_2}
\frac{\tilde{g}_k(-p_1)\tilde{g}^*_k(p_2):\pi_{p_1}\pi_{p_2}:_a}{2}=F+G\nonumber
\eea
where
\beq
F=\pin{k}\lpin{p_1}\lpin{p_2}
\frac{\tilde{g}_k(-p_1)\tilde{g}^*_k(p_2)\pi_{p_1}\pi_{p_2}}{2}=\pin{k}\frac{\pi^\L_{k}\pi^\L_{-k}}{2} \label{feq}
\eeq
and
\bea
G&=&-\pin{k}\lpin{p_1}\lpin{p_2}
\frac{\tilde{g}_k(-p_1)\tilde{g}^*_k(p_2)\omega_{p_2}2\pi\delta(p_1+p_2)}{4}\nonumber\\
&=&-\pin{k}\lpin{p} \tilde{g}_k(p)\tilde{g}^*_k(p) \frac{\omega_{p}}{4}.
\eea

In all, we see that $H\p_2$ is the sum of a scalar $E+G$ plus an operator
\beq
D+F=\frac{1}{2}\pin{k}\left[\pi^\L_k\pi^\L_{-k}+\omega_k^2\phi^\L_{k}\phi^\L_{-k}\right].\label{dfo}
\eeq
This looks like an infinite sum of quantum harmonic oscillators, however $\phi^\L_k$ and $\pi^\L_k$ are not quite canonical variables in the regulated theory as
\bea
i\beta_{k_1k_2}=[\phi^\L_{k_1},\pi^\L_{k_2}]&=&\lpin{p_1}\tilde{g}_{-{k_1}}(-p_1)\lpin{p_2}\tilde{g}_{-k_2}(-p_2)[\phi_{p_1},\pi_{p_2}]\label{beta}\\
&=&i\lpin{p}\tilde{g}_{-{k_1}}(p)\tilde{g}_{-k_2}(-p)=i\int dx\int dy\lpin{p}g_{-{k_1}}(x)g_{-k_2}(y)e^{ip(x-y)}\nonumber\\
&=&i\int dx\int dyg_{-{k_1}}(x)g_{-k_2}(y)\alpha(x-y).\nonumber
\eea
Nonetheless let us try to solve it as if it were a sum of harmonic oscillators, by defining
\bea
B_k^\ddag=\frac{\phi^\L_k}{2}-i\frac{\pi^\L_k}{2\omega_k}\hsp
B_{-k}=\omega_k\phi^\L_k+i\pi^\L_k
\eea
so that
\beq
\omega_k B_k^\ddag B_k=\frac{\omega^2_k\phi^\L_k\phi^\L_{-k}+\pi^\L_k\pi^\L_{-k}}{2}-\frac{\omega_k}{2}\lpin{p}\tilde{g}_{{-k}}(p)\tilde{g}_{k}(-p).
\eeq
Then we can rewrite the operator part of $H\p_2$ as 
\beq
D+F=I+J\hsp I=\pin{k}\omega_k B_k^\ddag B_k\hsp 
J=\pin{k}\lpin{p}\tilde{g}^*_{{k}}(-p)\tilde{g}_{k}(-p)\frac{\omega_k}{2} \label{j}
\eeq
The operators $B$ and $B^\ddag$ operators automatically solve analogous constraints to (\ref{ccons})
\beq
0= \pin{k}B^\ddag_k \tilde{g}_k(q)= \pin{k}B_k \tilde{g}_k(q)\hsp |q|>\Lambda.  \label{bcons}
\eeq

Summarizing, we may write $H\p_2$ as the sum of a scalar $E+G+J$ plus a term $I$ which is of the quantum harmonic oscillator form
\beq
H\p_2=-\pin{k}\lpin{p} \tilde{g}_k(p)\tilde{g}^*_k(p) \frac{\left(\omega_{p}-\omega_k\right)^2}{4\omega_p}+\pin{k}\omega_k B_k^\ddag B_k. \label{h2f}
\eeq
We have argued that at one loop this Hamiltonian completely characterizes the kink sector.   In \ref{irapp} we argue that the naive IR divergence in the $c$-number term at $k=p$ vanishes.

If we now set the regulator to infinity, then substituting (\ref{alim}) into (\ref{beta}) one finds the standard canonical commutation relations
\beq
\stackrel{\rm{lim}}{{}_{\Lambda\rightarrow\infty}}[\phi^\L_{k_1},\pi^\L_{k_2}]=i\int dx g_{-k_1}(x)g_{-k_2}(x)=i\delta(k_1+k_2).
\eeq
We can then easily read off the exact spectrum of the regularized Hamiltonian at one loop.  The leading order kink ground state $\vac_0$ is the state annihilated by all $B_k$ and it has mass given by the first term in (\ref{h2f}), which, when $\Lambda\rightarrow\infty$ indeed agrees with the formula in Refs.~\cite{cahill76,mekink,memassa}. 

Using
\bea
[B_{k_1},B_{k_2}]&=&(\omega_{k_2}-\omega_{k_1})\beta_{-k_1-k_2}\hsp
[B^\ddag_{k_1},B^\ddag_{k_2}]=\frac{\omega_{k_2}-\omega_{k_1}}{4\omega_{k_1}\omega_{k_2}}\beta_{k_1k_2}
\nonumber\\
{[B_{k_1},B^\ddag_{k_2}]}
&=&\frac{\omega_{k_1}+\omega_{k_2}}{2\omega_{k_2}}\beta_{-k_1k_2} \label{balg}
\eea
one finds
%\beq
%I B^\ddag_{k_2}\vac=\frac{1}{2}\pin{k_1} \frac{\omega_{k_1}}{\omega_{k_2}}(\omega_{k_1}+\omega_{k_2})\beta_{-k_1k_2} B^\ddag_{k_1}\vac.
%\eeq
%Therefore the excited states are generated by the eigenvectors of
%\beq
%I_{k_1 k_2}=\frac{1}{2} \frac{\omega_{k_1}}{\omega_{k_2}}(\omega_{k_1}+\omega_{k_2})\beta_{-k_1k_2}
%\eeq
%contracted with $B^\ddag_{k_1}$ and each of these eigenvectors increases the kink energy by the corresponding eigenvalue.  Note that only eigenvectors satisfying the constraint (\ref{bcons}) yield states as only these combinations of $B^\ddag_k$ are part of our operator algebra.  The constraint is preserved by the Hamiltonian and so the subspace satisfying the constraint is spanned by a subset of such eigenvectors.  
when $\Lambda\rightarrow\infty$, that
%\beq
%\beta_{-k_1k_2}\rightarrow\delta(k_1-k_2)\hsp
%I_{k_1k_2}=\omega_{k_1}\delta(k_1-k_2)
%\eeq
%and so 
the excited states can be obtained by acting with $B^\ddag_k$, each of which increases the energy by
\beq
\Delta E=\omega_k.
\eeq

Our compact notation, in which integrals over $k$ included sums over discrete states, hid the role of the zero mode.  In the case of the zero mode, $\omega_B=0$ and so the definition of $B^\ddag_B$ is singular.  However in that case the oscillator (\ref{dfo}) has no $\phi^2$ term and so its contribution to $H\p_2$ is simply the nonrelativistic center of mass kinetic energy $\pi^2_B/2$.  Thus the spectrum therefore also includes various center of mass momenta for the kink.  This, together wth the shape and the continuum quantum harmonic oscillator spectrum, yield the known one-loop spectrum of Ref.~\cite{dhn2}.

\subsection{One-Loop Kink Ground State of the Regularized Hamiltonian}

The nondiagonal nature of (\ref{balg}) means that the normal mode basis does not diagonalize $H\p_2$ when $\Lambda$ is finite.  Nevertheless, we will show that the ground state is still annihilated by all $B_{k}$ and hence, even at finite $\Lambda$, the one-loop kink mass is given by the first term in \eqref{h2f}.

Let us go back a few steps to (\ref{deq}) and (\ref{feq})
\beq
D+F=\pin{k}\lpin{p_1}\lpin{p_2} \frac{\tilde{g}_k(-p_1)\tilde{g}^*_k(p_2)}{2}\left(\pi_{p_1}\pi_{p_2}+\omega_{k}^2 \phi_{p_1}\phi_{p_2}
\right).\label{dfp}
\eeq
Define the eigenstates of the operators $\phi_p$ as
\beq
\phi_p|\psi\rangle=\psi_p|\psi\rangle
\eeq
where $\psi$ is a real-valued function on the interval $[-\Lambda,\Lambda]$.  Let us write the arbitrary state $|\Psi\rangle$  in the Schrodinger representation
\beq
|\Psi\rangle=\int \mathcal{D}\psi \Psi[\psi]|\psi\rangle
\eeq
where the integral is over all functions $\psi$ and $\Psi[\psi]$ is the Schrodinger wave functional.  As the eigenstates $|\psi\rangle$ are a complete basis, the state $|\Psi\rangle$ is arbitrary.  It follows that
\beq
\pi_p |\psi \rangle = - 2\pi i \int \mathcal{D}\psi \frac{\delta \Psi[\psi]}{\delta \psi_{-p}} |\psi \rangle~. \label{pi}
\eeq

Inserting the Ansatz
\beq
\Psi[\psi]=\exp{-\frac{1}{2}\lpin{p_1}\lpin{p_2}A_{p_1p_2}\psi_{p_1}\psi_{p_2}} \label{ans}
\eeq
into (\ref{dfp}) one finds an expression of the form
\beq
(D+F)|\Psi\rangle=\left(\mathcal{E}+\lpin{q_1}\lpin{q_2}B_{q_1q_2}\psi_{q_1}\psi_{q_2}\right)|\Psi\rangle
\eeq
where
\bea
B_{q_1q_2}&=&\frac{1}{2}\lpin{p_1}\lpin{p_2}\pin{k}\tilde{g}_k(p_1)\tilde{g}_{-k}(p_2)\left[-A_{p_1,-q_1}A_{p_2,-q_2}+\omega_k^2\delta(p_1-q_1)\delta(p_2-q_2)\right]\nonumber\\
&=&-\frac{1}{2}\lpin{p_1}\lpin{p_2}\pin{k}\tilde{g}_k(p_1)\tilde{g}_{-k}(p_2)A_{p_1,-q_1}A_{p_2,-q_2}+\frac{1}{2}\pin{k}\tilde{g}_k(q_1)\tilde{g}_{-k}(q_2)\omega_k^2\nonumber \\
\mathcal{E}&=&\frac{1}{2}\lpin{p_1}\lpin{p_2}\pin{k}\tilde{g}_k(p_1)\tilde{g}_{-k}(p_2)A_{-p_1,-p_2} ~.
\eea
Our strategy will be to find the $A_{p_1p_2}$ such that $B_{q_1q_2}=0$ and use it to find $\mathcal{E}$.  A sufficient condition for $B$ to vanish is
\beq \label{sufficient}
\lpin{p}\tilde{g}_k(p)A_{p,-q}=\omega_k\tilde{g}_k(q)
\eeq
for all $k$ and $q$. Multiplying by $\tilde{g}_{-k}(-r)$ and integrating over $k$, using completeness of $g$, one finds
\beq
\pin{k}\tilde{g}_{-k}(-r)\lpin{p}\tilde{g}_k(p)A_{p,-q}=A_{r,-q}=\pin{k}\tilde{g}_{-k}(-r)\omega_k\tilde{g}_k(q).
\eeq

We claim that this $A_{pq}$, inserted into the Ansatz (\ref{ans}), is the kink ground state, or more precisely $\df^\dag|K\rangle$.  To see this, note that it is annihilated by all $B_{k}$:
\bea
B_{-k} |\Psi \rangle &=& \int_{-\Lambda}^{\Lambda} \frac{dp}{2\pi} \tilde{g}_{-k}(-p) (\omega_k \phi_p + i \pi_p ) |\Psi \rangle \cr
& = & \int_{-\Lambda}^{\Lambda} \frac{dp}{2\pi} \tilde{g}_{-k}(-p) \left( \omega_k \psi_p - 2\pi \int_{-\Lambda}^{\Lambda} \frac{dq}{2\pi} A_{-p,q} \psi_q \right) |\Psi \rangle  \cr
&=& 0,
\eea
where the last step follows from \eqref{sufficient}.  Hence $|\Psi\rangle$ is annihilated by the last term of \eqref{h2f} and since this term is a positive operator $|\Psi\rangle$ must be the lowest energy state.  As a check, note that the corresponding energy $\mathcal{E}$ is
\bea
\nonumber
\mathcal{E}&=&-\frac{1}{2}\lpin{p_1}\lpin{p_2}\pink{2}\tilde{g}_{k_1}(p_1)\tilde{g}_{-k_1}(p_2)\tilde{g}_{-k_2}(p_1)\omega_{k_2}\tilde{g}_{k_2}(p_2)\nonumber\\
&=&\frac{1}{2}\lpin{p_1}\lpin{p_2}\pin{k_2}2\pi\delta(p_1+p_2)\tilde{g}_{-k_2}(p_1)\omega_{k_2}\tilde{g}_{k_2}(p_2)\nonumber\\
&=&\frac{1}{2}\lpin{p}\pin{k}\omega_k\tilde{g}_{-k}(-p)\tilde{g}_{k}(p),
\eea
which matches the $J$ from Eq.~(\ref{j}).

Therefore, even in the regularized theory, $J$ is the energy contribution from $D+F$, or more precisely the eigenvalue of the operator $D+F$ acting on the kink ground state.  Including the scalar terms in $H\p_2$, one finds that the total ground state energy of the regularized kink is
\beq
E+G+J=-\pin{k}\lpin{p} \tilde{g}_k(p)\tilde{g}^*_k(p) \frac{\left(\omega_{p}-\omega_k\right)^2}{4\omega_p}.
\eeq
This is the ground state energy for the unregularized kink found in Ref.~\cite{cahill76} using mode matching, but now with the $p$ integral cut off.  Therefore the limit $\Lambda\rightarrow\infty$ agrees with the known result.

%{\bf{One $k$ in $\mathcal{E}$ above should be a $-k$}}

\section{Remarks}

In this paper we have described a quantum model which is regularized by a cutoff from the beginning.  It exhibits a quantum kink and we found, at one-loop, its ground state and mass.  The results were almost trivial generalizations of the corresponding results in the unregularized theory, in which one merely restricts the domain of integration of the momentum $p$ to the interval $[-\Lambda,\Lambda]$.  This is in part because at one-loop, the theory is free, although it is diagonal in the normal mode basis $k$ and not the momentum $p$ basis.  However this was also suggested by the fact that two-dimensional scalar models can be  rendered finite by normal ordering without ever regularizing, and so all quantities may be computed without recourse to regularization \cite{rajaraman,mekink}.  Thus one may already suspect that no sign of regularization may remain when the regulator is taken to infinity, as it could have been avoided from the beginning.

In more interesting models, with fermions or more dimensions, normal ordering is not sufficient to remove all divergences.  Thus it is possible that the limit in which the regulator is taken to infinity will leave some nonzero residue, indeed in the case of supersymmetric kinks one may expect to arrive at the contribution from the one-loop anomaly \cite{anom}.

In the supersymmetric case one may hope that sufficient supersymmetry will allow a nonperturbative approach.  However we saw here that the kink solution itself has corrections of order $e^{-\Lambda/m}$.  In perturbation theory, we expect this to be always multiplied by finite powers of $\Lambda$ and so such contributions will vanish in the $\Lambda\rightarrow\infty$ limit, but in the nonperturbative regime, which is the relevant regime for applications to paradigms \cite{thooftconf,mandelconf,greensite} of QCD confinement, it is possible that these corrections will be physically relevant if not dominant.

\appendix

\section{Finding the Shift: The $\phi^4$ Double Well} \label{appOne}

In this appendix we evaluate the leading correction to $f^\L(x)$ in the case of the $\phi^4$ double well.  We show that it is exponentially small in $\Lambda/m$ but nonzero.  Here we consider the $\phi^4$ double well without shifting $\tilde{\phi}^\L$ by $v$ as in Eq.~(\ref{shift}).  Recall that, in momentum space, this corresponds a shift of the transformed field by $2\pi v\delta(p)$ and in particular it only affects the momentum space field at $p=0$.  

In this model the definition \eqref{fdef} is
\begin{align}
\tilde{V}[gf] =&~ \frac{1}{4}\left(g^2f^{2}-m^2/2\right)^2\nonumber\\
\tilde{V}\p[gf]/g =&~ g^2f^3-m^2f/2\nonumber\\
0 =&~ \int dx \left[-\partial^2_xf^{\L}(x)-m^2 f^\L(x)/2+g^2f^{\L 3}(x)\right]e^{ipx}, \qquad  |p|\leq \Lambda. \label{f4def}
\end{align}
The basic approach is to insert the finite-$\Lambda$ Fourier transform
\begin{align}\label{fLdashint}
f^\L(x) =&~ \dashint_{-\Lambda}^{\Lambda} \frac{dq}{2\pi} \tilde{f}(q) e^{- i q x}~, 
\end{align}
into \eqref{f4def}.  Our Ansatz for $\tilde{f}$ is the series expansion
\begin{equation}\label{ftildepert}
\tilde{f}(q) = \sum_{n = 0} \tilde{f}_n(q)~,
\end{equation}
where $\tilde{f}_0(q)$ is just the Fourier transform of the classical solution $f(x)$ and $\tilde{f}_{n\geq 1}(q)$ is bounded, for all $q$, by a polynomial in $\Lambda$ times $e^{-\pi n \Lambda/m}$.  We will then solve for the $\tilde{f}_n(q)$ perturbatively, and along the way will see that our Ansatz is consistent.  This will imply the main result of this appendix, that $f(x)$ can be expanded in a power series with the leading $\Lambda$ dependence suppressed by order $O(e^{-\pi \Lambda/m})$.%\blu{To make the claim/series well-defined I think we need to tell them that $\tilde{f}_0$ is, at least on any domain where it is bounded, the usual Fourier transform of the classical solution.}   We will see in the following that $\tilde{f}_n(q)$ is $O(e^{-2\pi n \Lambda/m})$ for fixed $q$, $|q| < \Lambda$.  \blu{Is this right?  I don't remember so well, but I thought there was a $q$-dependence in the exponential, coming from the $sinh$ term, so the best you can say is that it is bounded by order $O(e^{-\pi n \Lambda/m})$, as in the text under (\ref{finale})?  But I could well be remembering wrong.} At each order, $\tilde{f}_n(q)$ will take the form of a polynomial in $\Lambda$ multiplying this exponentially damped behavior.  We will construct $\tilde{f}_1$ explicitly and see that it is bounded by a quantity of $O(\Lambda \cdot e^{-\pi \Lambda/m})$ for all $|q| \leq \Lambda$.

There are, however, two technical points that require explanation before proceeding.  First, we must define the $\dashint$ symbol in \eqref{fLdashint}.  Ordinary integration, for $f(x)$ a classical kink solution, would be ill-defined at $q=0$.  We are free to define the Fourier transform $\tilde{f}$ as we like, so long as we are able to use it to demonstrate the main result written above.  Therefore, we choose $\dashint$ to be a principal value integral, defined by 
\begin{equation}\label{PVint}
\dashint_{-\Lambda}^{\Lambda} := \lim_{\epsilon \to 0_+} \left( \int_{-\Lambda}^{-\epsilon} + \int_{\epsilon}^{\Lambda} \right)~.
\end{equation}
This type of integral can be used to obtain the standard kink profile $f^{(\infty)}(x) = \frac{m}{2g} \tanh(mx/2)$ from its Fourier transform $\tilde{f}_0(q) = \frac{\sqrt{2} \pi i}{g} \csch(\pi q/m)$:
\begin{equation}\label{tildef0}
\frac{m}{2g} \tanh(mx/2) = \dashint_{-\infty}^{\infty} \frac{dq}{2\pi} \left( \frac{\sqrt{2} \pi i}{g} \csch(\pi q/m) \right) e^{-i q x}~,
\end{equation}
The integral on the right would be undefined without the principle value prescription.  More generally, as a distribution the Fourier transform $\mathcal{F}[f^{(\infty)}]$ acts on any test function $b(q)$ via integration against $\tilde{f}_0(q)$ with the principle value prescription.  Indeed, the proper setting for the kink profile and its Fourier transform is to view both as tempered distributions.  The kink profile, $f^{(\infty)}$, is a nonsingular distribution, meaning that it is defined globally by a smooth function.  The Fourier transform $\tilde{f}_0 = \mathcal{F}[f^{(\infty)}]$ is singular, meaning that it can only be represented locally by a smooth function (the csch above); its definition as a distribution contains more information -- namely the principal value prescription.  Although we do not a priori know the explicit finite-$\Lambda$ analogs, $f^{(\Lambda)}$ and $\tilde{f}$, we we expect they have the same large-$|x|$ asymptotics and the same singularity structure at $q = 0$ as the leading order configurations $f^{(\infty)}, \tilde{f}_0$, respectively, hence the appearance of the principal value integral in \eqref{fLdashint}.

A second technical point is the following.  The standard result for smooth functions that the Fourier transform of the pointwise product is the convolution of Fourier transforms,
\begin{equation}\label{convolutionthm}
\mathcal{F}[f \cdot g] = \frac{1}{2\pi} \mathcal{F}[f] \star \mathcal{F}[g]~,
\end{equation}
also holds for tempered distributions,\footnote{It holds for distributions when the product is defined.  In general this involves a condition on the wave front sets of the distributions \cite{hormander}.  In the case of $f^{(\infty)}$, which is defined globally by integration against a smooth function, the wave front set is trivial.  The product of this distribution with any other temprered distribution is defined and the convolution theorem holds.} where the factor of $2\pi$ is due to our Fourier transform conventions.  Here, if $f,g$ are ordinary functions then $(f \cdot g)(x) = f(x) g(x)$ denotes the pointwise product and $(f \star g)(x) = \int dy f(y) g(x-y)$ denotes the convolution.  The convolution of two distributions is defined through the convolution of a distribution with a smooth test function as follows.  If $f$ is a distribution defined locally by $f(x)$ and $b$ a smooth function with compact support, then the convolution $f \star b$ is a smooth function with value $(f \star b)(x) = \int dy f(y) b(x-y)$.  The convolution of two distributions $f,g,$ is then the unique distribution such that $(f \star g) \star b = f \star (g \star b)$.  This definition and the corresponding convolution theorem can be extended to
\begin{equation}
\mathcal{F}[f_1 \cdot  \ldots \cdot f_n] = \frac{1}{(2\pi)^{n-1}} \tilde{f}_1 \star \ldots \star \tilde{f}_n~,
\end{equation}
where $\tilde{f}_j = \mathcal{F}[f_j]$ and the distribution on the right is defined through sequential action on a test function.

An instructive example is to show that \eqref{convolutionthm} holds for the kink profile $f = g = f^{(\infty)}$ and its Fourier transform.  Using $\tanh^2 = 1 - \sech^2$, one finds that $\mathcal{F}[f^{(\infty)} \cdot f^{(\infty)}]$ has
\begin{equation}\label{fsqFT}
\mathcal{F}[f^{(\infty)}\cdot f^{(\infty)}](q) = \frac{\pi m^2}{g^2} \delta(q) - \frac{2\pi q}{g^2 \sinh(\pi q/m)} ~.
\end{equation}
Below, we will recover this result in the $\Lambda \to \infty$ limit of a finite-$\Lambda$ computation of the convolution, $\tilde{f}_0 \star \tilde{f}_0$. 

We can think of $f^{(\Lambda)}$ in \eqref{fLdashint} as the ordinary inverse Fourier transform of $\tilde{f}$ if we define $\tilde{f}$ to vanish outside of $[-\Lambda,\Lambda]$.  For example, if $\tilde{f}_0$, initially defined on all of $\mathbb{R}$, is set to zero outside $[-\Lambda,\Lambda]$, the resulting distribution is still a tempered distribution (now with compact support).  Henceforth we will write $\tilde{f}^{(\Lambda)}$ for distributions with support on $[-\Lambda,\Lambda]$.  In particular, for any test function $b$ we have
\begin{equation}
\dashint_{-\Lambda}^{\Lambda} \frac{dq}{2\pi} \tilde{f}_0(q) b(q) = \dashint_{-\infty}^{\infty} \frac{dq}{2\pi} \tilde{f}_{0}^{(\Lambda)}(q) b(q)~.
\end{equation}
The convolution theorem continues to hold for such distributions, so that for powers of $f^{(\Lambda)}$ we have
\begin{align}\label{Lconvolutionthm}
\mathcal{F}[(f^{(\Lambda)})^n] =&~ \frac{1}{(2\pi)^{n-1}} \tilde{f}^{(\Lambda)} \star \ldots \star \tilde{f}^{(\Lambda)}~.
\end{align}
Note this means that the support of the $n$-$\star$ convolution must be $[-n\Lambda,n\Lambda]$ such that
\begin{align}
& \int_{-n \Lambda}^{n\Lambda} \left(\tilde{f}^{(\Lambda)} \star \ldots \star \tilde{f}^{(\Lambda)}\right)(q) b(p-q) =  \dashint_{-\Lambda}^{\Lambda} dq_1 \tilde{f}(q_1) \dashint_{-\Lambda}^{\Lambda} dq_2 \tilde{f}(q_2) \cdots \dashint_{-\Lambda}^{\Lambda} dq_n \tilde{f}(q_n) b(p - \Sigma_i q_i) ~. 
\end{align}
This reflects the fact that the pointwise product of $n$ $f^{(\Lambda)}$'s on the left of \eqref{Lconvolutionthm} contains momentum modes up to $|q| = n \Lambda$.

With these preliminaries out of the way we can proceed with the perturbative solution of \eqref{f4def}.  Inserting \eqref{fLdashint} into \eqref{f4def} gives 
\begin{align}\label{integraleqn}
0 =&~ \left( p^2 - \frac{m^2}{2} \right) \tilde{f}^{(\Lambda)}(p) + \frac{g^2}{(2\pi)^2} \left(\tilde{f}^{(\Lambda)} \star \tilde{f}^{(\Lambda)} \star \tilde{f}^{(\Lambda)}\right)(p) ~, \qquad |p| \leq \Lambda~.
\end{align}
This is an integral equation for the local function $\tilde{f}^{(\Lambda)}(p)$ on $[-\Lambda,\Lambda]$.  We expect the distribution $\tilde{f}^{(\Lambda)}$ to be singular at $p = 0$ and defined via the principal value prescription as discussed above.  Therefore, we restrict attention in \eqref{integraleqn} to $p \neq 0$.

Inserting the expansion \eqref{ftildepert} into \eqref{integraleqn}, and assuming that $\tilde{f}_0$ solves the equation when $\Lambda \to \infty$ (which we will check), we find the following linear equation for the first correction, $\tilde{f}_1$:
\begin{equation}\label{linearizedftildeeqn}
\left( \left( p^2 - \tfrac{m^2}{2} \right) \cdot \tilde{f}_{1}^{(\Lambda)} + \frac{3 g^2}{(2\pi)^2} \left(\tilde{f}_{0}^{(\Lambda)} \star \tilde{f}_{0}^{(\Lambda)}\right) \star \tilde{f}_{1}^{(\Lambda)} \right)(p) = F_1(p)~, \qquad |p| < \Lambda~,
\end{equation}
where $F_1(p)$ is the first finite-$\Lambda$ correction from the non-linear term evaluated on the leading-order solution:
\begin{equation}\label{f1source}
F_1(p) := - \frac{g^2}{(2\pi)^2} \left(\tilde{f}_{0}^{(\Lambda)} \star \tilde{f}_{0}^{(\Lambda)} \star \tilde{f}_{0}^{(\Lambda)}\right)(p) \bigg|_{\textrm{first subleading behavior in $m/\Lambda$}} ~.
\end{equation}
Hence the main tasks now are to compute the finite-$\Lambda$ convolutions of $\tilde{f}_{0}^{(\Lambda)}$ with itself appearing in \eqref{linearizedftildeeqn}, \eqref{f1source}, and to invert the linear operator in \eqref{linearizedftildeeqn}.  In the following we describe the approach and present the results but suppress all intermediate steps.\footnote{A more detailed presentation of this calculation may appear elsewhere.}  

\begin{figure}
\begin{center}
\includegraphics[scale=0.5]{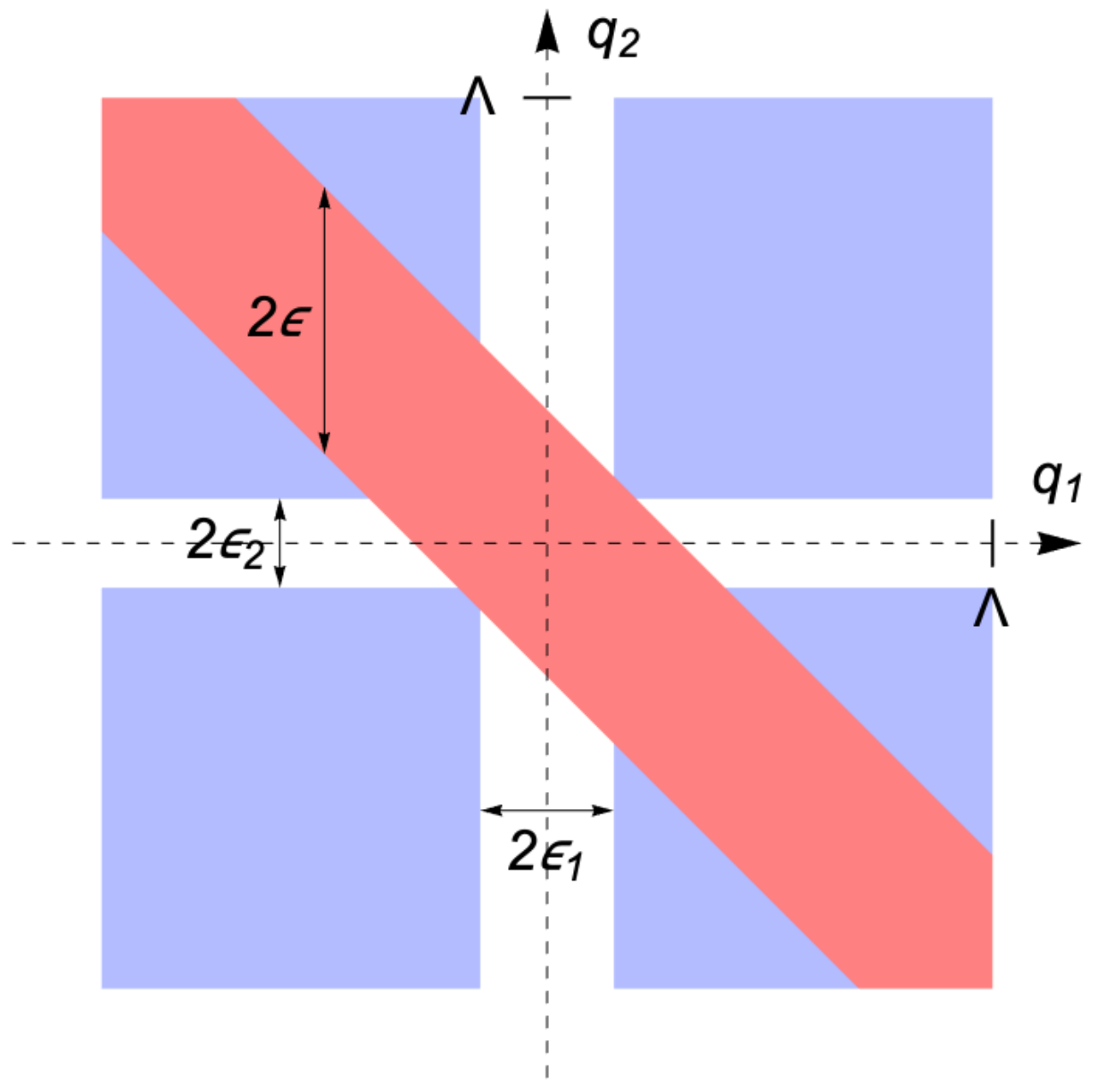}
\end{center}
\caption{Integration region for the right-hand side of \eqref{f0starf0}.  For the integral over the blue region we change variables from $(q_1,q_2)$ to $(q,q_2)$ where $q = q_1 + q_2$.  The limits of integration on $q_2$ then become a function of $q,\Lambda,\epsilon_{1,2}$.  For example, when $\epsilon < q < \Lambda - \epsilon_1$, the integral over $q_2$ covers $[q-\Lambda,-\epsilon_2] \cup [\epsilon_2, q - \epsilon_1] \cup [q + \epsilon_1, \Lambda]$.  The test function does not depend on $q_2$ and can be pulled out of the $q_2$ integral.  For all cases $\epsilon < |q| < 2\Lambda$, the result of the $q_2$ integration, after taking the limits $\epsilon_2 \to 0$ followed by $\epsilon_1 \to 0$, is captured by the second term on the right-hand side of \eqref{f0starf0result}.  For the integral over the pink region we instead expand the test function around $q = 0$ and compute leading term in the resulting integrand.  That computation requires first integrating over $q_2$ for fixed $q_1$ (where the limits of integration depend on $q_1,\epsilon,\epsilon_2$), then taking the $\epsilon_2 \to 0$ limit, then integrating over $q_1$, and finally taking $\epsilon \to 0$.}
\label{appAfig}
\end{figure}

The first convolution we need is $\frac{1}{2\pi} (\tilde{f}_{0}^{(\Lambda)} \star \tilde{f}_{0}^{(\Lambda)})$, given by
\begin{align}\label{f0starf0}
& \frac{1}{2\pi} \int_{-2\Lambda}^{2\Lambda} dq \left(\tilde{f}_{0}^{(\Lambda)} \star \tilde{f}_{0}^{(\Lambda)}\right)(q) b(p - q) = \cr
& \qquad \qquad =  -\frac{\pi}{g^2} \dashint_{-\Lambda}^{\Lambda} dq_1 \csch(\pi q_1/m) \dashint_{-\Lambda}^{\Lambda} dq_2 \csch(\pi q_2/m) b(p - q_1 - q_2)~,
\end{align}
for any smooth test function $b$.  We introduce infinitesimal parameters $\epsilon_{1,2}$ associated with the principal value integrals over $q_{1,2}$ respectively, and we can assume $\epsilon_2 < \epsilon_1$ since the $\epsilon_2 \to 0$ limit is to be taken first.  The basic idea is to change integration variables in the double integral from $(q_1,q_2)$ to $(q,q_2)$, where $q = q_1 + q_2$, and carry out the $q_2$ integral to extract $(\tilde{f}_{0}^{(\Lambda)} \star \tilde{f}_{0}^{(\Lambda)})(q)$.  This, however, is only valid away from $q = 0$, since it involves exchanging the $\epsilon_2 \to 0$ limit with the integral over $q_1$.  If $q = 0$ then the $\csch(\pi q_1/m)$ factor is not smooth at the $q_2$ pole, and one must take $\epsilon_2 \to 0$ limit first, before evaluating the $q_1$ integral.  Therefore we divide the integration into two pieces: one over a region $R_\epsilon = \{ (q_1,q_2) \in [-\Lambda,\Lambda]^2 : |q_1 + q_2| < \epsilon \}$ and the other over $[-\Lambda, \Lambda]^2 \setminus R_\epsilon$.  We take $\epsilon > \epsilon_1 + \epsilon_2$ and send $\epsilon \to 0$ at the very end.  See Figure \ref{appAfig}.

%%%%%%%%%
%%%%%%%%%

The integral over $R_\epsilon$ gives a result for the right-hand side of \eqref{f0starf0} that is proportional to $b(p)$ in the limit $\epsilon \to 0$, and hence this corresponds to a delta function contribution to $(\tilde{f}_{0}^{(\Lambda)} \star \tilde{f}_{0}^{(\Lambda)})(q)$.  The coefficient can be evaluated and turns out to be $\Lambda$-independent, and this term matches the $\delta$-function term in \eqref{fsqFT}.  Meanwhile the integral over $[-\Lambda, \Lambda]^2 \setminus R_\epsilon$ gives a $\Lambda$-dependent contribution.  In total we find
\begin{align}\label{f0starf0result}
\frac{1}{2\pi} \left(\tilde{f}_{0}^{(\Lambda)} \star \tilde{f}_{0}^{(\Lambda)}\right)(q) =&~  \frac{\pi m^2}{g^2} \delta(q) - \frac{2m}{g^2 \sinh(\pi |q|/m)} \ln \left[ \frac{\sinh(\pi \Lambda/m)}{\sinh(\pi |\Lambda - |q| |/m)} \right]~.
\end{align}
This agrees with the right-hand side of \eqref{fsqFT} in the $\Lambda \to \infty$ limit for fixed $q$, as required by the convolution theorem.  It has integrable logarithmic singularities at $|q| = \Lambda$.

For the double convolution we use
\begin{align}\label{f0cubedaction}
\int_{-3\Lambda}^{3\Lambda}dp \left(\tilde{f}_{0}^{(\Lambda)} \star \tilde{f}_{0}^{(\Lambda)} \star \tilde{f}_{0}^{(\Lambda)}\right)(p) b(k - p) =&~  \dashint_{-\Lambda}^{\Lambda} dq_1 \tilde{f}_0(q_1) \dashint_{-\Lambda}^{\Lambda} \tilde{f}_0(q_2) \dashint_{-\Lambda}^{\Lambda} dq_3 \tilde{f}_0(q_3) b(k- \Sigma_i q_i) \cr
=&~  \dashint_{-\Lambda}^{\Lambda} dq_1 \tilde{f}_0(q_1) \int_{-2\Lambda}^{2\Lambda} dq (\tilde{f}_{0}^{(\Lambda)} \star \tilde{f}_{0}^{(\Lambda)})(q) b(k - q_1 - q) ~.
\end{align}
We apply \eqref{f0starf0} for $(\tilde{f}_0 \star \tilde{f}_0)(q)$ and change variables in the remaining double integral from $(q_1,q)$ to $(p,q)$ where $p = q_1 + q$.  Integrating over $q$ will then allow us to extract the double convolution $(\tilde{f}_{0}^{(\Lambda)} \star \tilde{f}_{0}^{(\Lambda)} \star \tilde{f}_{0}^{(\Lambda)})(p)$.  Note there is no principal value prescription required for the distribution $(\tilde{f}_{0}^{(\Lambda)} \star \tilde{f}_{0}^{(\Lambda)})$ as it has no poles.  Thus for all $p \neq 0$ it is possible to obtain a local representation of the double convolution by carrying out the $q$ integral only.   Furthermore, although the support of the double convolution is $[-3\Lambda,3\Lambda]$, we only require it in the range $[-\Lambda,\Lambda]$ for the solution to \eqref{integraleqn}.  For any $p$ in this range, the integration over $q$ covers $[p-\Lambda,p-\epsilon_1] \cup [p + \epsilon_1, p + \Lambda]$.  Here the outer limits are due to $|q_1| \leq \Lambda$ and the inner limits are due to the principal value prescription on the $q_1$ integration.  Hence, we have that
\begin{align}\label{f0starf0starf0}
\notag
\frac{1}{(2\pi)^2} \left(\tilde{f}_{0}^{(\Lambda)} \star \tilde{f}_{0}^{(\Lambda)} \star \tilde{f}^{(\Lambda)}\right)(p) =&~ \frac{1}{(2\pi)^2} \lim_{\epsilon_1 \to 0} \left( \int_{p - \Lambda}^{p-\epsilon_1} + \int_{p+\epsilon_1}^{p+\Lambda} \right) dq \tilde{f}_0(p-q) (\tilde{f}_0 \star_{\Lambda} \tilde{f}_0)(q) \\
=&~ \frac{m^2}{2 g^2} \tilde{f}_0(p) - \frac{1}{g^2} F(p)~, \qquad |p| < \Lambda~,
\end{align}
where the first term arises from the Dirac delta term in \eqref{f0starf0} and we've defined
\begin{align}\label{Fofp}
F(p) :=&~  \frac{(2m) \sqrt{2} \pi i}{(2\pi) g} \lim_{\epsilon_1 \to 0} \left( \int_{p - \Lambda}^{p-\epsilon_1} + \int_{p+\epsilon_1}^{p+\Lambda} \right) dq ~ \csch(\pi (p-q)/m) \csch(\pi |q|/m) \times \cr
&~ \qquad \qquad \qquad \qquad \qquad \qquad \qquad \qquad \quad  \times  \ln \left[ \frac{\sinh(\pi \Lambda/m)}{\sinh(\pi |\Lambda - |q||/m)} \right] ~,
\end{align}
as giving the contribution from the remaining piece of \eqref{f0starf0}.  We've pulled out an explicit factor of $-1/g^2$ from the definition of $F$ so that $F_1(p)$ in \eqref{f1source} is precisely the first subleading term (with respect to the exponential behavior) in the large $\Lambda$ expansion of $F(p)$.

By changing variables $q \to -q$, one sees that the second integral in \eqref{Fofp} is the negative of the first with $p \to -p$.  Thus $F(p)$ is an odd function of $p$.  The integral can be evaluated in closed form in terms of logarithms and dilogarithms.  We record the full result here since it is important in demonstrating our claim that the perturbative expansion is an expansion in $e^{-2\pi \Lambda/m}$.  Setting $\varepsilon = e^{-\pi \Lambda/m}$ and $z = e^{\pi p/m}$ for shorthand, we find
\begin{align}
F(p) =&~  \frac{i \sqrt{2} m^2}{\pi g ~ \sinh(\pi p/m)} \bigg\{ \frac{\pi^2 p^2}{m^2} + \frac{\pi p}{m} \ln \left[ \frac{1 - \varepsilon^2 z^{-2}}{1 - \varepsilon^2 z^2} \right] + \frac{3 \pi \Lambda}{m} \ln \left[ \frac{ (1 - \varepsilon^2 z^2)(1 - \varepsilon^2 z^{-2})}{ (1-\varepsilon^2)^2} \right] + \cr
&~ + \ln \left[ \frac{ (z-z^{-1})^2}{(1-\varepsilon^2)^2} \right] \ln \left[ \frac{ (1-\varepsilon^2)^2}{(1-\varepsilon^2 z^2)(1 - \varepsilon^2 z^{-2})} \right] + \cr
&~ + \ln \left[ \frac{| z^2 - 1| }{1 - \varepsilon^2 z^2} \right] \ln \left[ \frac{1- \varepsilon^2}{1 - \varepsilon^2 z^2} \right]  + \ln \left[ \frac{| z^{-2} - 1| }{1 - \varepsilon^2 z^{-2}} \right] \ln \left[ \frac{1- \varepsilon^2}{1 - \varepsilon^2 z^{-2}} \right] + \cr
&~ - \frac{1}{2} \litwo(\varepsilon^2 z^2) - \frac{1}{2} \litwo( \varepsilon^2 z^{-2}) + \litwo(\varepsilon^2) - \litwo\left( \frac{ 1 - z^2}{1 - \varepsilon^2} \right)  - \litwo\left( \frac{ 1 - z^{-2}}{1 - \varepsilon^2} \right) + \cr
&~ + \litwo\left( \frac{1 - z^2}{1 - \varepsilon^2 z^2} \right)  + \litwo\left( \frac{1 - z^{-2}}{1 - \varepsilon^2 z^{-2}} \right) + \litwo \left( \frac{\varepsilon^2 (1- z^2)}{(1- \varepsilon^2 z^2)} \right) + \litwo \left( \frac{\varepsilon^2 (1- z^{-2})}{(1- \varepsilon^2 z^{-2})} \right)  \bigg\} ~,
\end{align}
and we have checked this result against numerical integration.  Noting that $\litwo(x) = x + O(x^2)$ for small $x$, it is easy to see that $F(p)$ has a power series expansion in $\varepsilon^2$, where every term after the very first one in the curly brackets above begins at $O(\varepsilon^2)$:
\begin{equation}
F(p) = p^2 \tilde{f}_0(p) + \sum_{n=1}^{\infty} F_{n}(p) ~,
\end{equation}
where, for fixed $p$, $F_n = O(e^{-2\pi n \Lambda/m})$ times a linear function in $\Lambda$.  Inserting this result back into the double convolution \eqref{f0starf0starf0} gives
\begin{equation}
\left(\tilde{f}_{0}^{(\Lambda)} \star \tilde{f}_{0}^{(\Lambda)} \star \tilde{f}_{0}^{(\Lambda)}\right)(p) = \frac{1}{g^2} \left( \frac{m^2}{2} - p^2 \right) \tilde{f}_0(p) - \frac{1}{g^2} \sum_{n=1}^{\infty} F_{n}(p) ~.
\end{equation}
Therefore $\tilde{f}_0(p)$ is indeed the leading order solution to the integral equation, \eqref{integraleqn}.  Furthermore, we find that the source term in the equation for the first correction, \eqref{linearizedftildeeqn}, is
\begin{equation}\label{F1source}
F_1(p) = -\frac{6 i \sqrt{2} m^2}{\pi g} \left\{ \frac{2\pi \Lambda}{m} + 1 - \ln [4 \sinh^2(\pi p/m)] \right\} \sinh(\pi p/m) e^{-2\pi \Lambda/m} ~.
\end{equation}

Notice that, while for fixed $p$ the source term \eqref{F1source} is $O(\Lambda \cdot e^{-2\pi \Lambda/m})$, it is bounded by a quantity of $O(\Lambda \cdot e^{-\pi \Lambda/m})$ for all $p \in [-\Lambda,\Lambda]$.  In particular $F_1$ vanishes at $p=0$ as do all of the $F_n$.

Now consider the integral operator on the left-hand side of the linearized equation \eqref{linearizedftildeeqn}, which we denote $\tilde{\Delta}^{(\Lambda)}(p-q)$:
\begin{align}
\dashint_{-\Lambda}^{\Lambda} dq \tilde{\Delta}^{(\Lambda)}(p-q) \tilde{f}_{1}^{(\Lambda)}(q) =&~  \left( \left( p^2 - \tfrac{m^2}{2} \right) \cdot \tilde{f}_{1}^{(\Lambda)} + \frac{3 g^2}{(2\pi)^2} \left(\tilde{f}_{0}^{(\Lambda)} \star \tilde{f}_{0}^{(\Lambda)}\right) \star \tilde{f}_{1}^{(\Lambda)} \right)(p) ~.
\end{align}
The double convolution can be written in the same form as \eqref{f0cubedaction}, but with $\tilde{f}_{0}^{(\Lambda)}(q_1)$ replaced with $\tilde{f}_{1}^{(\Lambda)}(q_1)$.  The principal value integration from $[-\Lambda,\Lambda]$ arises from changing variables in the analog of \eqref{f0starf0starf0} from $q$ to $q' = p-q$.  Thus
\begin{align}\label{DeltaL}
\quad \tilde{\Delta}^{(\Lambda)}(p-q) =&~ \left( p^2 - \tfrac{m^2}{2} \right) \delta(p-q) + \frac{3 g^2}{(2\pi)^2} \left(\tilde{f}_{0}^{(\Lambda)} \star \tilde{f}_{0}^{(\Lambda)}\right)(p-q) \cr
=&~ (p^2 + m^2) \delta(p-q) - \frac{3m}{\pi \sinh(\frac{\pi |p-q|}{m})} \ln \left[ \frac{\sinh(\frac{\pi \Lambda}{m})}{\sinh(\frac{\pi |\Lambda - |p-q||}{m})} \right]~ . \qquad
\end{align}
We would like to find an inverse for $\tilde{\Delta}^{(\Lambda)}$, whose kernel is a Green function that we denote $\tilde{G}^{(\Lambda)}(q,k)$, for $q,k \in [-\Lambda,\Lambda]$.  In order to extract the leading behavior of $\tilde{f}_{1}^{(\Lambda)}$, however, it is sufficient to find an approximate inverse, whose kernel we denote $\tilde{G}(q,k)$, such that
\begin{equation}\label{approxinverse}
\int_{-\Lambda}^{\Lambda} dq \left( \int_{-\Lambda}^{\Lambda} dq' \tilde{\Delta}^{(\Lambda)}(p-q') \tilde{G}(q',q) \right) F_{n}(q) = F_{n}(p) (1 +  O((m/\Lambda)^2) ) ~,
\end{equation}
where we recall that $F_n$ is bounded, odd, and $O(\Lambda \cdot e^{-2\pi n \Lambda/m})$ for fixed $p$ with $|p| < \Lambda$.

A standard Neumann series for the inverse, based on the form of \eqref{DeltaL} as a diagonal operator plus ``correction,'' where the correction is the second term, may not provide a sufficiently good approximation demonstrate this, since the correction is $O(1)$ in a neighborhood of the diagonal.  A natural ansatz for a better first approximation to $\tilde{G}^{(\Lambda)}$ is based on its $\Lambda \to \infty$ counterpart, which we describe next.

We claim that the approximate Green function can be taken to be the Green function of the standard $\Lambda \to \infty$ fluctuation operator, $\tilde{\Delta}^{(\infty)}$, which we denote by $\tilde{G}^{(\infty)}(p,q)$, restricted to $p,q \in [-\Lambda,\Lambda]$.  We will discuss the error in this approximation after presenting $\tilde{\Delta}^{(\infty)}$ and its Green function.  We have
\begin{align}\label{tildedeltainf}
\quad \tilde{\Delta}^{(\infty)}(p-q) =&~  (p^2 + m^2) \delta(p-q) - \frac{3(p-q)}{\sinh(\pi (p-q)/m)} ~ ,
\end{align}
which can be obtained from the Fourier transform of the position space fluctuation operator around the kink:
\begin{align}
\tilde{\Delta}^{(\infty)}(p-q) =&~ \int dx dy e^{i p x - i q y} \Delta^{(\infty)}(x-y)~, \qquad \textrm{with} \cr
\Delta^{(\infty)}(x-y) =&~ \delta(x-y) \left( - \partial_{x}^2 - \frac{m^2}{2} + 3 g^2 f^{(\infty)}(x)^2 \right)~,
\end{align}
for which the eigenmodes are the $g_k(x)$ in \eqref{fluceqn}.  $\Delta^{(\infty)}$ does not have an inverse due to the zero-mode $g_B$.  It does have an inverse on the orthogonal complement of $g_B$, whose integral kernel is the Green function $G^{(\infty)}(x,y)$ satisfying
\begin{equation}\label{Greeneqn}
\int dy \Delta(x-y) G(y,z) = \delta(x-z) - g_B(x) g_B(z)~.
\end{equation}
Imposing orthogonality to $g_B$ and exponential fall-off at large $|x-y|$, the solution to \eqref{Greeneqn} is
\begin{align}\label{positionGreen}
G^{(\infty)}(x,y) =&~ \frac{e^{2m x_<} + e^{-2 m x_>} + 8 (e^{m x_<} + e^{-m x_>}) - 6 m (x_> - x_<) - 8}{32 m~\cosh^2(m x_>/2) \cosh^2(m x_</2)} ~,
\end{align}
where $x_{>} = \max(x,y)$ and $x_{<} = \min(x,y)$.  

Then $\tilde{G}^{(\infty)}(p,q)$ is the Fourier transform satisfying
\begin{equation}\label{DeltaG1}
\int_{-\infty}^{\infty} dq' \tilde{\Delta}(p - q') \tilde{G}(q',q) = \delta(p-q) - \tilde{g}_B(p)\tilde{g}_B(-q) ~,
\end{equation}
and we can compute it explicitly:
\begin{align}\label{momentumGreen}
\tilde{G}^{(\infty)}(p,q) =&~ \frac{2\pi}{(q^2 + m^2)} \delta(p-q) + \tilde{G}_{\rm reg}(p,q)~, \qquad \textrm{with} \cr
\tilde{G}_{\rm reg}(p,q) =&~ \frac{\pi}{\sinh(\pi (p-q)/m)} \bigg\{ \frac{12 p}{(q^2 + m^2)^2} - \frac{12 q}{(p^2 + m^2)^2} - \frac{13 p}{m^2 (q^2 + m^2)} + \frac{13 q}{m^2 (p^2 + m^2)} + \cr
&~ \qquad \qquad \qquad \qquad  +\frac{6}{m^4} (p-q) + \frac{11 p q}{m^5} \left( \coth(\pi p/m) - \coth(\pi q/m) \right) + \cr
&~ \qquad \qquad \qquad \qquad  - \frac{6 pq}{m^5} ~ {\rm Im} \left( \psi^{(1)}(-1 + \tfrac{i p}{m}) - \psi^{(1)}(-1 + \tfrac{i q}{m}) \right) \bigg\}~.
\end{align}
where $\psi^{(1)}(z) = \frac{d}{dz} \psi(z)$ with $\psi(z)$ the digamma function.  Note that $\tilde{G}(p,q) = \tilde{G}(q,p)$.

The regular piece $\tilde{G}_{\rm reg}$ is smooth and bounded on all of $\mathbb{R}^2$.  The pole from the csch pre-factor along the diagonal is canceled by a zero from the quantity in curly brackets.  For fixed $p$, the large $|q|$ behavior of the quantity in curly brackets is linear and therefore $G_{\rm reg}(p,q) = O(|q| e^{-\pi |q|})$ as $|q| \to \infty$.  Furthermore, along the diagonal, one can show that $G_{\rm reg}(p,p)$ falls off like $|p|^{-4}$ for large $p$.

We use these facts to argue that the restriction of $\tilde{G}^{(\infty)}(p,q)$ to $p,q \in [-\Lambda,\Lambda]$ can be taken as the approximate inverse $\tilde{G}$ in \eqref{approxinverse}:
\begin{equation}
\tilde{G}(p,q) = \tilde{G}^{(\infty)}(p,q) \bigg|_{[-\Lambda,\Lambda]^2} ~.
\end{equation}
There are two sources of error in this approximation.  First there is the error in replacing the inverse of $\tilde{\Delta}^{(\Lambda)}$ by the inverse of $\tilde{\Delta}^{(\infty)} |_{[-\Lambda,\Lambda]^2}$.  Then there is the error in approximating the inverse of the latter operator, which we will think of as a restriction of $\tilde{\Delta}^{(\infty)}$ to an upper-left block, with the upper-left block of its inverse.\footnote{There is no issue regarding the fact that $\tilde{G}^{(\infty)}$ is only an inverse to $\tilde{\Delta}^{(\infty)}$ on the orthogonal complement to the zero-mode $\tilde{g}_B$.  The reason is that we can restrict interest to $\tilde{G}$ acting on functions $F_n(q)$ that are odd on $q \in [-\Lambda,\Lambda]$ and so are orthogonal to $\tilde{g}_{\rm B}$, which is an even function of $q$.}  We discuss each in turn.  

To simplify notation, let $T = \tilde{\Delta}^{(\Lambda)}$ be the integral operator we have, and let $T_0 = \tilde{\Delta}^{(\infty)} |_{[-\Lambda,\Lambda]^2}$.  Let the difference be $\delta T$ with kernel
\begin{equation}
\delta T(p,q) = \frac{3(p-q)}{\sinh(\pi (p-q)/m)} - \frac{3m}{\pi \sinh(\frac{\pi |p-q|}{m})} \ln \left[ \frac{\sinh(\frac{\pi \Lambda}{m})}{\sinh(\frac{\pi |\Lambda - |p-q||}{m})} \right] ~,
\end{equation}
which vanishes along the diagonal and is exponentially small in $m/\Lambda$ in a neighborhood of the diagonal.  Again, we note the logarithmic singularity at $|p-q| = \Lambda$ is integrable and has a coefficient that is exponentially small in $m/\Lambda$.  Thus we expect the corrections in approximating $T^{-1}$ by $T_{0}^{-1}$ from the Neemann series,
\begin{align}\label{TvsT0}
T^{-1} =&~ (T_0 + \delta T)^{-1} = T_{0}^{-1} - T_{0}^{-1} \delta T T_{0}^{-1} + T_{0}^{-1} \delta T T_{0}^{-1} \delta T T_{0}^{-1} - + \cdots \cr
\approx &~ T_{0}^{-1} ~,
\end{align}
to be exponentially small in $m/\Lambda$.

Now, $T_0$ is the restriction of $\mathcal{T} = \tilde{\Delta}^{(\infty)}$ to functions with support on $[-\Lambda,\Lambda]$, while we define $\tilde{G}$ as the analogous restriction of $\tilde{G}^{(\infty)}$, and what we know is that $\mathcal{T} \tilde{G}^{(\infty)} = \mathbf{1}$.  A finite-dimensional matrix analog of our problem is that we have that
\begin{equation}
\mathcal{T} = \left( \begin{array}{c c} T_0 & T_{12} \\ T_{21} & T_{22} \end{array} \right) ~, \qquad \tilde{G}^{(\infty)} = \left( \begin{array}{c c} \tilde{G} & \tilde{G}_{12} \\ \tilde{G}_{21} & \tilde{G}_{22} \end{array} \right)~
\end{equation}
are inverses of each other and we would like an expression for $T_{0}^{-1}$.  Assuming $G_{22}$ is invertible, then the finite-dimensional formula from block inversion is
\begin{equation}\label{T0toG}
T_{0}^{-1} = \tilde{G} - \tilde{G}_{12} \tilde{G}_{22}^{-1} \tilde{G}_{21}~.
\end{equation}

We assume the same formula holds in our setting, and we expect $\tilde{G}_{22}$ to be invertible since it has the form of a diagonal matrix plus a small correction.  Then the above remarks on the properties of $\tilde{G}^{(\infty)}$ imply that $\tilde{G}_{12}(p,q)$, with $|p| < \Lambda$ and $|q| > \Lambda$, is generally exponentially suppressed, except in neighborhoods of the corners where $|p| \sim \Lambda \sim |q|$.  In these neighborhoods we still have that $\tilde{G}_{12}(p,q)$ is bounded by a quantity of $O(|q|^{-4})$.  Analogous remarks apply to $\tilde{G}_{21} = (\tilde{G}_{12})^T$.  Therefore we expect that $\tilde{G}_{12} \tilde{G}_{22}^{-1} \tilde{G}_{21}$ is suppressed\footnote{In more detail, when $p,q$ are away from the diagonal $\tilde{G}_{12} \tilde{G}_{22}^{-1} \tilde{G}_{21}$ has double the exponential suppression in the distance from the diagonal as $\tilde{G}$.  Along the diagonal, assuming the dominant behavior of $(\tilde{G}_{22})^{-1}(p,q)$ is $\sim p^2 \delta(p-q)$, the contribution of $\tilde{G}_{12} \tilde{G}_{22}^{-1} \tilde{G}_{21}$ is estimated by an integral of the form $\int_{0}^{\infty} dx (x + \frac{\Lambda}{m})^{-6} e^{-x}$, which is $O(1/\Lambda^6)$.} by at least $O(m^2/\Lambda^2)$ relative to $\tilde{G}$ for any $p,q \in [-\Lambda,\Lambda]$.  Therefore combining \eqref{TvsT0} and \eqref{T0toG} we write
\begin{equation}
\tilde{G}^{(\Lambda)}(p,q) \equiv T^{-1}(p,q) = \tilde{G}(p,q) \left( 1 + O(m^2/\Lambda^2) \right)~,
\end{equation}
which gives \eqref{approxinverse}.

It follows that the solution to \eqref{linearizedftildeeqn} for the first correction is
\begin{align}\label{f1intsol}
\tilde{f}_1(p) = \int_{-\Lambda}^{\Lambda} dq \tilde{G}(p,q) F_1(q) \times (1 + O(m^2/\Lambda^2)) ~.
\end{align}
Note that for fixed $p$ the exponential damping in $|p-q|$ from the csch pre-factor of $\tilde{G}_{\rm reg}$ balances the exponential growth from the sinh pre-factor of $F_1$.  Since the integrand then grows linearly in $q$ at large $q$ and the explicit leading $\Lambda$ dependence of $F_1$ is $O(\Lambda \cdot e^{-2\pi \Lambda/m})$, one sees that the leading behavior of $\tilde{f}_1$ will be $O(\Lambda^3 \cdot e^{-2\pi \Lambda/m})$.  Since the $\Lambda$ dependence of the integrand of \eqref{f1intsol} is simple, the leading $\Lambda$ behavior of the integral can be computed by applying the fundamental theorem of calculus.  The result is
\begin{align}
& \tilde{f}_{1}^{(\Lambda)}(p,\Lambda) = \cr
&=  - \frac{4\pi i \sqrt{2}}{g} \sinh(\pi p/m) \left\{ \frac{ (6 p^4 - m^2 p^2 + 5 m^4)}{(p^2 + m^2)^2} + \frac{3 p}{2m} \operatorname{Im}\left[ \psi^{(1)}(-1+i \tfrac{p}{m}) \right] \right\} \frac{\Lambda^3}{m^3} e^{-2\pi \Lambda/m}  + \cr
& \qquad \quad + O\left( \tfrac{\Lambda^2}{m^2} e^{-2\pi \Lambda/m}\right) ~. \label{finale}
\end{align}
Thus for fixed $p$ we have $\tilde{f}_{1}^{(\Lambda)}= O(\Lambda^3 \cdot e^{-2\pi \Lambda/m})$, and for all $p \in [-\Lambda, \Lambda]$ we have that $\tilde{f}_{1}^{(\Lambda)}$ is bounded by a quantity of $O(\Lambda^3 \cdot e^{-\pi \Lambda/m})$.

Since $\tilde{f}_{1}^{(\Lambda)}$ has the same exponentially suppressed form as the source $F_1$, it follows from the above analysis that the linearized equation for the $n^{\rm th}$ correction, $\tilde{f}_{n}^{(\Lambda)}$, will be of the form $\tilde{\Delta}^{(\Lambda)} \cdot \tilde{f}_{n}^{(\Lambda)} = F_{n}'$, where $F_{n}'$ has the same suppression as $F_n$.  Hence $\tilde{f}_{n}^{(\Lambda)}$ will also have the same exponential suppression as $F_n$, and will generally involve a polynomial in $\Lambda/m$ whose degree grows with $n$.

\section{IR Divergences} \label{irapp}

At two or three loops, depending on the theory in general the energy of the kink ground state is IR divergent.  More specifically, the Sine-Gordon kink has a divergence at three loops, the $\phi^4$ kink at two, and if the potential expanded around a minimum corresponding to each end of the kink begins its polynomial expansion at $\phi^n$, the first IR divergence will occur at $n-1$ loops.  These IR divergences do not affect the kink mass as they also appear, at one less loop, in the vacuum energy and the kink mass is the difference between the two energies.

Here at one loop we found the energy
\beq
Q_1=-\pin{k}\lpin{p} \tilde{g}_k(p)\tilde{g}^*_k(p) \frac{\left(\omega_{p}-\omega_k\right)^2}{4\omega_p}.
\eeq
In the cases of the Sine-Gordon and $\phi^4$ modes, $\tilde{g}_k(p)$ contains a $\delta(k-p)$ term.  One of these $\delta$ functions may be eliminated by the integration over $k$, but the other potentially leaves a divergence already at one loop.  In coordinate space, it is easily seen that the origin of this divergence is a constant energy density, and so it is an IR divergence.  

Such IR divergences have been noted since the first papers \cite{dhn2} on kink mass calculations, and are usually regularized by compactifying the space using periodic boundary conditions.  This is quite a high price to pay\footnote{While the price is high, the cut-off compactified theory has a finite number of modes and so can be treated nonperturbatively using Monte Carlo \cite{lee,slava,bajnok} and variational \cite{tensor20} techniques.}, as the kink itself does not satisfy periodic boundary conditions.  One may instead impose boundary conditions that are satisfied by the kink, but then they will not be satisfied by the vacuum.  

We will now argue that, at least in the present case, the constant energy density is exactly zero and so this messy issue may be avoided.  First, let us return to position space
\beq
Q_1=-\lpin{p}\pin{k}\int dx\int dy {g}_{-k}(x)  {g}_k(y) e^{ip(y-x)} \frac{\left(\omega_{p}-\omega_k\right)^2}{4\omega_p}.
\eeq
The potential divergence now arises from the plane wave terms
\beq
g_k(x)\supset \frac{|k|}{\omega_k}e^{-ikx}.
\eeq
Thus the potentially divergent term in $Q_1$ is
\beq
\hat{Q}_1=-\lpin{p}\pin{k}\int dx\int dy e^{i(p-k)(y-x)} \frac{k^2}{\omega^2_k}\frac{\left(\omega_{p}-\omega_k\right)^2}{4\omega_p}.
\eeq
The potential divergence arises from $p\sim k$ so let us expand in $1/\epsilon$ where $\epsilon=p-k$.  At leading order
\beq
\omega_p-\omega_k=\frac{\epsilon}{\omega_p}
\eeq
and so at leading order, at $p\sim k$, we have
\beq
\hat{Q}_1=-\pin{p}\frac{p^2}{4\omega_p^5} \pin{\epsilon}\int dx\int dy e^{i\epsilon(y-x)} \epsilon^2.
\eeq

Notice that the exponential is periodic in both $x$ and $y$ with period $2\pi/\epsilon$.  It is also bounded by $1$, as is its norm.  Therefore integrating $e^{i\epsilon(y-x)}$ over any rectangle on the $x-y$ plane, the integral will never exceed $4\pi^2/\epsilon^2$.  Thus the integral of $\epsilon^2 e^{i\epsilon(y-x)}$ never exceeds $4\pi^2$ and in particular is bounded.  When $\epsilon\neq 0$ the integral vanishes in the sense of a distribution as it is periodic.  Therefore, after integrating over $x$ and $y$ one arrives at a function of $\epsilon$ which is bounded and vanishes except on the measure zero set $\epsilon=0$.  The integral over $\epsilon$ is therefore equal to zero.  Thus, at leading order in $1/\epsilon$, this integral vanishes.  We conclude that there is no small $\epsilon$ divergence, and so the potential IR divergence is not present.  

Note that the above derivation did not use the $p$ integral, it was performed independently at each value of $p$.  Therefore it is not affected by the cutoff at $\Lambda$.

\section* {Acknowledgement}

\noindent
JE is supported by the CAS Key Research Program of Frontier Sciences grant QYZDY-SSW-SLH006 and the NSFC MianShang grants 11875296 and 11675223.   JE also thanks the Recruitment Program of High-end Foreign Experts for support.  ABR is supported by NSF grant number PHY-2112781.

\end{document}